\documentclass[a4paper,11pt]{article}
\pdfoutput=1 

\usepackage{jheppub} 

\usepackage[T1]{fontenc} 

\title{\boldmath Lepton sourced baryon asymmetry in the fourth generation model}

\author[a,1]{Hsiang-nan Li\note{Corresponding author.}}

\affiliation[a]{Institute of Physics, Academia Sinica,\\Taipei, Taiwan 115, Republic of China}

\emailAdd{hnli@phys.sinica.edu.tw}

\abstract{We demonstrate that the observed baryon asymmetry in the Universe can be accommodated 
in the extended Standard Model with sequential fourth generation fermions (SM4). We first 
construct the dimension-6 effective operators of the type
$-i(\Phi^\dagger\Phi)\bar F_L\Phi f_R$ induced by fourth generation quarks, which 
carry the $CP$ violation (CPV) source from the $4\times 4$ Cabibbo-Kobayashi-Maskawa (CKM) 
matrix, $\Phi$ ($F_L$, $f_R$) being a Higgs double (left-handed fermion doublet, 
right-handed fermion singlet). The required inputs of the fourth generation fermion 
masses were derived in our previous dispersive analyses on heavy quark decays and neutral 
meson mixing. The similar framework allows the determination of 
the $4\times 4$ CKM matrix elements $V_{ib'}$, $i=u$, $c$ and $t$, such that the strength 
of the CPV source can be evaluated unambiguously. The dimension-6 operators associated with 
fourth generation leptons, as implemented into the formalism for the electroweak baryogenesis 
in the literature, lead to the baryon-over-entropy ratio $\eta_B\approx 10^{-10}$.}

\begin{document} 
\maketitle
\flushbottom

\section{Introduction}
\label{sec1}

The dispersive analyses on the flavor structure of the Standard Model (SM)
\cite{Li:2023dqi,Li:2023yay,Li:2023ncg,Li:2024awx} have suggested that the mass hierarchy 
and the distinct mixing patters of quarks and leptons are governed by the analyticity of 
SM dynamics. Our series of works echoes the ``$S$-matrix bootstrap conjecture" advocated by 
Geoffrey Chew in 1960s \cite{Chew:1962mpd} that a well-defined but infinite set of 
self-consistency conditions (based on analyticity, unitarity, causality, etc.) determines 
uniquely the aspects of particles in nature \cite{Cushing:1985zz,vanLeeuwen:2024uzj}. 
The bootstrap conjecture was proposed for strong interaction originally, and then extended 
to electroweak interaction \cite{Chew:1982qy}. Motivated by the dynamical interpretation of 
the SM flavor structure, we explored the sequential fourth generation model as the  
most economical extension of the SM (SM4), to which no free parameters are added 
\cite{Li:2023fim,Li:2024xnl}. For example, the mass $m_{t'}\approx 200$ TeV 
($m_{b'}= 2.7$ TeV) of a fourth generation quark $t'$ ($b'$) was demanded by the 
dispersion relations for the mixing between the neutral quark states $t'\bar u$ and 
$\bar t' u$ ($b'\bar d$ and $\bar b' d$) \cite{Li:2023fim}. The multiple intermediate 
channels involved in the responsible box diagrams, i.e., the $db'$, $sb'$ and $bb'$ 
($ut$ and $ct$) channels in the $t'$ ($b'$) case, give consistent results. The 
mass $m_{\tau'}=270$ GeV ($m_4=170$ GeV) of a fourth generation charged (neutral) 
lepton $\tau'$ ($\nu'$) was predicted by investigating the dispersion relation for the 
decay $\tau'\to \nu\bar t d$ ($t\to d e^+\nu'$), $\nu$ and $e^+$ being a light neutrino 
and a positron, respectively \cite{Li:2024xnl}.

We have demonstrated that the heavy quark condensate $\langle\bar t't'+\bar b'b'\rangle<0$ 
is established as the Yukawa coupling exceeds the critical value $g_Q^c\approx 9.1$ 
\cite{Li:2025bon}. The heavy quark condensate, producing a quadratic term $\mu^2\phi^2/2$ 
with the mass parameter $\mu^2<0$ in the Higgs potential, breaks the electroweak symmetry
\cite{Hung:2009hy,Enkhbat:2011vp}. Fourth generation leptons with the smaller Yukawa couplings 
of $O(1)$ do not constitute the condensate. Additional heavy scalars (or pseudoscalars) then 
appear as bound states \cite{Bardeen:1989ds} of $t'$ and $b'$ quarks with the masses above 
a TeV scale in a Yukawa potential (the Yukawa coupling $9.1$ corresponds to a quark mass 
1.6 TeV). The contributions from $\bar b'b'$ scalars to the Higgs boson production via gluon 
fusion and to the Higgs decay into a photon pair were found to be of $O(10^{-3})$ and 
$O(10^{-2})$ of the top quark one \cite{Li:2023fim}, respectively. Other heavier scalars 
contribute even less. These estimates elucidated why superheavy fourth generation quarks 
bypass the experimental constraints from Higgs boson production and decay 
\cite{Chen:2012wz,Eberhardt:2012gv,Djouadi:2012ae,Kuflik:2012ai}. If fourth 
generation quarks appear in the loops associated with electroweak precision observables, whose 
contributions are not described effectively by bound states, they must come from charged 
currents. Their effects are then strongly suppressed by the small $4\times 4$ CKM matrix 
elements, whose values will be delivered in Sec.~\ref{sec3}.
Fourth generation leptons, with the masses of the electroweak scale, are too light to form 
bound states. We evaluated the contribution from charged leptons to the $H\to \gamma\gamma$ 
decay width $\Gamma_{\gamma\gamma}$, and evinced that its effect is within the current 
uncertainties of $\Gamma_{\gamma\gamma}$ measurements. The impact on the oblique parameters 
from fourth generation quarks and leptons, based on the formulas in \cite{He:2001tp}, is also 
permitted by the experimental errors \cite{Li:2024xnl}. The search for heavy leptons at 
colliders with the mass 
of the electroweak scale was simulated in \cite{Marcano:2024bjs,Joshi:2025yls}. 

It has been known that the baryon asymmetry in the Universe (BAU) requires three 
necessary conditions \cite{Sakharov:1967dj}: baryon number violation, $C$ and $CP$ violation, 
and departure from thermal equilibrium. Whether the BAU can be understood in the SM was 
inspected in the literature 
\cite{Farrar:1993sp,Farrar:1993hn,Gavela:1993ts,Gavela:1994ds,Gavela:1994dt}. 
The SM provides the baryon number violating sphaleron interaction, the $CP$ violation (CPV)
source from the Kobayashi-Maskawa mechanism, and the electroweak phase transition 
(EWPT), which is, however, not strongly first-order 
\cite{Kajantie:1996mn,Rummukainen:1998as,Csikor:1998eu,Aoki:1999fi}. 
The extra heavy scalars in the SM4 modify the Higgs potential through the one-loop 
Coleman-Weinberg correction $V_1(\phi)$ at zero temperature \cite{Coleman:1973jx} 
and the one-loop temperature-dependent correction $V_T(\phi)$ \cite{Dolan:1973qd}, such that 
a barrier between the trivial vacuum and a nontrivial vacuum of the effective Higgs potential 
is built up. The resultant ratio $\phi_c/T_c \approx 0.9$ \cite{Li:2025bon}, where $\phi_c$ 
denotes the location of the nontrivial vacuum at the critical temperature $T_c$, roughly 
meets the criterion for the strongly first-order EWPT $\phi_c/T_c \gtrsim 1$. That is, 
fourth generation quarks play a crucial role through their bound states for achieving the 
first-order EWPT.

This work will present a quantitative verification that the aforementioned SM4 accounts for 
the observed BAU in terms of the baryon-over-entropy ratio $\eta_B\approx 10^{-10}$. The 
standard workflow for calculating the BAU within the electroweak baryogenesis (EWBG), from 
the derivation of semi-classical CPV sources in plasma particle collisions with expanding 
bubbles, to the formulation of transport equations for particle number densities, and then 
to the implementation of weak sphaleron effects, is referred to recent reviews 
\cite{vandeVis:2025efm,Barni:2025ifb}. It has 
been proposed \cite{Balazs:2016yvi,DeVries:2018aul} to test EWBG by means of effective 
CPV operators, instead of engaging cumbersome exercise on individual models explicitly. 
The dimension-6 effective operators of the type $-i(\Phi^\dagger\Phi)\bar F_L\Phi f_R$, 
$\Phi$ ($F_L$, $f_R$) being a Higgs double (left-handed fermion doublet, right-handed 
fermion singlet) were considered in \cite{Zhang:1994fb}. As noticed in 
\cite{DeVries:2018aul}, the BAU from the CPV quark operators is inefficient because of the 
washout by strong sphalerons, and the lepton sourced BAU turns out to dominate
\cite{Joyce:1994bi,Chung:2009cb,Chung:2008aya}. The coefficient $g_\tau/\Lambda_\tau^2$ was 
assumed for the $\tau$ lepton operator in \cite{DeVries:2018aul} with the $\tau$ lepton 
Yukawa coupling $g_\tau$ and the new physics scale $\Lambda_\tau=1$ TeV, which satisfies the 
experimental constraint from the electron electric dipole moment (EDM) \cite{Fuchs:2020uoc}. 
It was shown that the $\tau$-sourced BAU suffices to accommodate the observed $\eta_B$. 
However, a more recent study based on the assumption of Minimal Flavor Violation gave a 
negative conclusion \cite{Alonso-Gonzalez:2021jsa}, so whether the $\tau$-sourced scenario
is viable remains unsettled \cite{Li:2024mts,Liu:2025xxb} in a phenomenological viewpoint.

We will adopt the approach in \cite{DeVries:2018aul}, and predict the BAU in the SM4. 
The dimension-6 effective operators are constructed from fourth generation quarks 
unambiguously with all essential inputs being specified. Their coefficients depend on 
the fourth generation quark masses, which were obtained in our earlier dispersive analyses 
as stated before, and on the $4\times 4$ Cabibbo-Kobayashi-Maskawa (CKM) matrix elements, 
which carries new $CP$-odd phases. The latter will be deduced in a similar framework 
applied to the neutral quark state mixing. The coefficients are found to be proportional 
to $g_f/\Lambda^2$ with the Yukawa couplings $g_f$ for the fermions $f$ in the 
effective operators. The new physics scale $\Lambda\approx 16$ TeV, being universal for 
all flavors $f$, is much higher than $\Lambda_\tau=1$ TeV set in \cite{DeVries:2018aul}, 
implying that the $\tau$-sourced BAU is short by two orders of magnitude in our setup. It 
will be elaborated that the effective operators for fourth generation leptons $\tau'$ and 
$\nu'$, whose contributions are enhanced by the $O(1)$ Yukawa couplings, replace 
the $\tau$ operator in \cite{DeVries:2018aul}; the implementation 
of the $\tau'$ and $\nu'$ operators into the formalism for the EWBG 
\cite{DeVries:2018aul} yields the ratio $\eta_B\approx 10^{-10}$ in 
agreement with the observation.

The rest of the paper is organized as follows. The relevant dimension-6 effective operators 
are established in Sec.~\ref{sec2} by matching them to the diagrams consisting of fourth generation 
quarks in the full theory. It will be seen that the CPV source from the $4\times 4$ CKM 
matrix is introduced at the three-loop level, and the associated Jarlskog invariant appears 
naturally. The $4\times 4$ CKM matrix elements $V_{ub'}$, $V_{cb'}$ and $V_{tb'}$ are 
acquired in Sec.~\ref{sec3} by solving the dispersive constraints from the $c\bar u$-$\bar cu$, 
$t\bar u$-$\bar tu$ and $t\bar c$-$\bar tc$ neutral quark state mixings. Given the $4\times 4$ 
CKM matrix elements and the coefficients $g_f/\Lambda^2$ of the dimension-6 operators, we 
extract the prediction for the BAU straightforwardly from the results in \cite{DeVries:2018aul} 
in Sec.~\ref{sec4}. Section~\ref{sec5} contains the conclusion: fourth generation quarks make the 
first-order EWPT via the interactions between their bound-state scalars and Higgs bosons, 
and fourth generation leptons create the BAU through their CPV collisions with emergent bubbles 
during the first-order EWPT. Appendix~\ref{appA} collects the details of attaining the 
dimension-6 effective operators with the CPV sources. We take this opportunity to reanalyze 
the $4\times 4$ Pontecorvo–Maki–Nakagawa–Sakata (PMNS) matrix for the lepton mixing  
in Appendix~\ref{appB}, confirming the outcomes for the involved mixing angles and $CP$ phase in 
our previous publication \cite{Li:2024xnl}.

\section{Effective operators}
\label{sec2}

It has been recognized that new physics is requested to achieve the first-order EWPT and 
to enhance CPV, which are the necessary conditions for realizing the BAU. An effective 
theory, allowing a model-independent and systematic investigations of relevant 
new physics impacts, has been probed intensively in the literature 
\cite{Bodeker:2004ws,Fromme:2006wx,Beneito:2025fzf,ThomasArun:2025rgx,ThomasArun:2025dav}; 
an effective $\phi^6/(8M^2)$ term, 
$M$ being a new physics scale, was introduced into the Higgs potential 
to prompt the first-order EWPT. As to the CPV, the dimension-6 operators in the form 
$-i(\Phi^\dagger\Phi)\bar F_L\Phi f_R$ with an imaginary coefficient were 
studied \cite{Kosnik:2025srw}, whose strength was constrained by available experimental 
data for the electron EDM. The new CPV sources were then included into the transport 
equations that describe SM particle and bubble interactions. The particle number 
densities were solved from the transport equations, and the BAU from, e.g., the top quark 
source \cite{Balazs:2016yvi}, was calculated. We 
take advantage of the above formalism, constructing the dimension-6
effective operators with definite coefficients based on our SM4 setup and reading off the 
predictions for the BAU that correspond to the strength of the effective operators. This 
strategy saves the tedious handling of transport processes and sphaleron effects 
\cite{Lee:2004we} specific to the SM4. 

\begin{figure}[tbp]
\centering 
\includegraphics[scale=0.9]{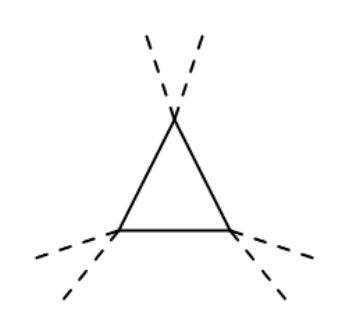}

\caption{\label{fig1}
Diagram for producing the dimension-6 effective operator $\phi^6/(8M^2)$, where a dashed
(solid) line represents a Higgs boson (heavy scalar).}
\end{figure}

The aforementioned effective operators appear in certain renormalizable extensions 
of the SM, such as models with vector-like quarks and Higgs triplets \cite{Huang:2015izx}. 
We illustrate that fourth generation quarks also produce the $\phi^6/(8M^2)$ term with the 
desired strength. This is expected, since our SM4 setup has been shown to induce the 
first-order EWPT \cite{Li:2025bon}. Firstly, the Higgs-heavy-quark interaction could be 
replaced by the effective interaction $\lambda'\phi^2\eta^2/2$ in the symmetry broken phase, 
where the coupling takes a value $\lambda'\approx 3.8$ \cite{Li:2025bon} and $\eta$ denotes 
a bound-state scalar formed by fourth generation quarks. We then derive the dimension-6 
operator by integrating out the $\eta$ field at the electroweak scale. The triangular 
diagram in Fig.~\ref{fig1} with six external Higgs legs leads to the effective vertex
\begin{eqnarray}
\frac{-i}{8M^2}=
8\frac{(2\lambda')^3}{3!}\int\frac{d^4l}{(2\pi)^4}
\left(\frac{1}{l^2-m_\eta^2}\right)^3,\label{8M}
\end{eqnarray} 
where the coefficient $8$ is from the combinatorial factor, $2\lambda'$ is associated 
with the $\phi^2\eta^2$ vertex, and $1/3!$ comes with the expansion to the third power of 
the interaction term. The momenta of external Higgs bosons, being of order of the 
electroweak scale and much lower than the heavy scalar mass $m_\eta\approx 3.2$ TeV 
\cite{Li:2023fim}, have been neglected.

A simple computation of the integral in Eq.~(\ref{8M}) gives
\begin{eqnarray}
M=\sqrt{\frac{3}{8\lambda^{\prime 3}}}\pi m_\eta\approx 830\;{\rm GeV}.
\end{eqnarray}
Referring to Fig.~1 in \cite{Bodeker:2004ws}, we see that the point $(m_H,M)=(125, 830)$ 
in units of GeV is near the the curve labeled by $\xi\equiv \phi_c/T_c=1$. This observation 
coincides with the one in \cite{Li:2025bon} that our SM4 setup generates the first-order 
EWPT characterized by $\xi$ around unity. The consistency encourages our attempt to 
translate the SM4 into the effective theory in \cite{Bodeker:2004ws,Fromme:2006wx}. 
We emphasize that the derivation of the effective operator $\phi^6/(8M^2)$ 
is not for constructing the effective Higgs potential $V_{\rm eff}$, which is utilized
to examine the EWPT. The effective potential $V_{\rm eff}$ has been constructed in 
\cite{Li:2025bon} according to the standard procedure, which sums the one-loop corrections 
resulting from the effective interaction $\lambda'\phi^2\eta^2/2$ in the broken phase to 
all powers in $\phi$. The heavy quark contributions to $V_{\rm eff}$ are thus
effectively encoded in the bound-state description in a reliable way.

It seems to be claimed in \cite{Kikukawa:2009mu} that the EWPT in a model with
fourth generation fermions is not strongly first-order. There is crucial distinction between 
our setup and the one in \cite{Kikukawa:2009mu}. The latter starts with strong four-fermion 
interactions among the four generation, and establishes the effective theory below the 
compositeness scale $\Lambda_{4f}$. Their Higgs bosons are composite, formed by fourth 
generation fermions, so compositeness conditions need to be specified. Whether the 
first-order EWPT is achievable sensitively depends on choices of compositeness conditions. 
Their claim that the SM4 does not lead to the strongly first-order EWPT was made for the 
compositeness scale $\Lambda_{4f}\lesssim 2.3$ TeV. However, the compositeness condition 
at $\Lambda_{4f}=3.7$ TeV, classified as Criterion B in \cite{Kikukawa:2009mu}, can. Because 
a Higgs boson is not a composite of fourth generation quarks in our setup, the analysis does 
not suffer the ambiguity from compositeness conditions. It was then shown that our SM4  
realizes the strongly first-order EWPT. Hence, there is no contradiction between our 
conclusion and the one in \cite{Kikukawa:2009mu}.

The dimension-6 effective operator $-is_fg_f(\Phi^\dagger\Phi)\bar F_L\Phi f_R/\Lambda^2$ 
results in, after the electroweak symmetry breaking,
\begin{eqnarray}
-i\frac{s_fg_fv^2}{\sqrt{2}\Lambda^2}\phi\bar f\gamma_5 f, \label{hff}
\end{eqnarray}
with the Yukawa coupling $g_f$ of a fermion $f$ \cite{deVries:2017ncy,DeVries:2018aul} and
the vacuum expectation value (VEV) $v$ of a Higgs field. The discretionary sign $s_f=+1$ or 
$-1$ was chosen to accommodate the observed BAU \cite{DeVries:2018aul}, e.g., $s_t=+1$ 
for a top quark and $s_\tau=-1$ for a $\tau$ lepton. The new physics 
scale $\Lambda$ was constrained by the data of the electron EDM, which could vary with the 
flavor $f$, depending on the sensitivity of the operator to the data. We will 
construct the operator in Eq.~(\ref{hff}) from the SM4 below. It will be highlighted 
that both the sign $s_f$ and the new physics scale $\Lambda$, i.e., the strength 
of the operator, are fixed unambiguously by matching Eq.~(\ref{hff}) to the full theory 
with fourth generation quarks.

\begin{figure}[tbp]
\centering 
\includegraphics[scale=0.7]{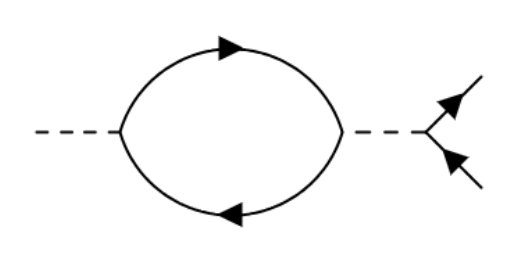}\hspace{1.0cm}
\includegraphics[scale=0.7]{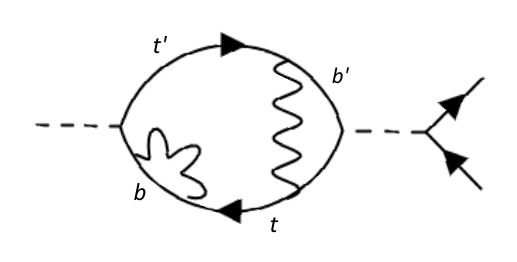}

(a)\hspace{7.0cm}(b)
\caption{\label{fig2}
(a) One-loop diagram and (b) three-loop diagram for the effective operator 
$\phi\bar f\gamma_5 f$, where an arrowed solid line denotes a heavy quark and a wavy line 
denotes a charged scalar. }
\end{figure}

We explicate that the CPV source of the effective operator arises in three-loop diagrams 
with fourth generation quarks. It is more convenient to assess heavy quark contributions
in the $R_\xi$ gauge than in the unitary gauge. If the latter is employed, the 
effect from a weak boson exchange between heavy quarks will be governed by the $q_\mu q_\nu$ 
piece in a weak boson propagator with the momentum $q$. After applying equations of motion, 
we have the Yukawa couplings associated with fourth generation quarks, which are much larger 
than the weak coupling. This contribution is identical to that from the exchange of a 
charged scalar or a neutral pseudoscalar in the $R_\xi$ gauge. Figure~\ref{fig2}(a) displays 
a fourth generation quark loop, say, a $t'$ loop, with one vertex from a Higgs boson $H$ and 
another from a pseudoscalar $\phi_Z$ in a Higgs doublet. The virtual $\phi_Z$ splits into a 
fermion pair $f\bar f$, developing the pseudoscalar current $\bar f\gamma_5 f$. This diagram 
does not endow a $CP$-odd phase apparently. We must add charged scalars to Fig.~\ref{fig2}(a), 
whose attachments to the $t'$ quark loop bring in $4\times 4$ CKM matrix elements. 
Adding a single charged scalar does not make an imaginary contribution, for the associated 
product of the CKM matrix elements, like $V^*_{t'b'}V_{t'b'}$, is real. We need at least two 
charged scalars. 

To maximize the CPV source, the external Higgs boson attaches a $t'$ quark, which has
the largest mass $m_{t'}$, i.e., the largest Yukawa coupling $g_{t'}$ among quarks at the 
electroweak scale. One charged scalar crosses the pseudoscalar vertex, inducing a pseudoscalar 
penguin for the flavor-changing transition $t'\to t$ with an internal $b'$ quark. If the 
internal $b'$ quark is exchanged for a $b$ quark, the $CP$-odd phase will be down by the lower 
$b$ quark Yukawa coupling. To form a complete set of penguin diagrams, the self-energy 
correction from the charged scalar is included, which also involves the $t'\to t$ transition 
with an internal $b'$ quark. In this case, the pseudoscalar attaches the $t'$, instead of 
$t$, quark to keep the contribution maximal. For a similar reason, another charged scalar 
should form a self-energy correction with the $t\to t'$ transition, instead of crossing the 
Higgs boson vertex. Otherwise, the contribution will be suppressed at least by the $b$ quark 
Yukawa coupling. The inclusion of the two charged scalars thus gives rise to the three-loop 
diagram in Fig.~\ref{fig2}(b), which contains the sequence of flavor-changing transitions 
$t'\to b'\to t \to b\to t'$ with the product of the $4\times 4$ CKM matrix elements 
$V^*_{t'b'}V_{tb'}V^*_{tb}V_{t'b}$. The imaginary part of this product, serving 
precisely as a Jarlskog invariant \cite{Jarlskog:1985ht,Jarlskog:1985cw,Shaposhnikov:1986jp}, 
provides a CPV source. Substituting other lighter up-type (down-type) quarks for $t$ ($b$) 
yields minor contributions owing to smaller Yukawa couplings. 


We remark that the parametrization in Eq.~(\ref{hff}) with the coefficient being 
proportional to the Yukawa coupling $g_f$ \cite{DeVries:2018aul} applies to heavy 
fermions $f=t,\tau',\nu'$, and the $CP$-odd phase identified above is maximal. For light 
fermions, the parametrization does not hold in the SM4, because contributions from other 
types of diagrams may be comparable to that from Fig.~\ref{fig2}(b). For instance, 
changing the virtual pseudoscalar to a virtual $Z$ boson causes a suppression factor 
$g/g_{b'}$ from the vertex on the internal $b'$ quark line, $g$ being the weak coupling. 
At the same time, the change brings in an enhancement factor $g/g_f$ from the vertex 
on the light fermion line of $f$. The new diagram, generating an electroweak penguin with 
$V\pm A$ currents, will not be negligible, if $g^2$ and $g_{b'}g_f$ are of the similar order 
of magnitude. The contributions from light fermions to the BAU are expected to be 
insignificant \cite{DeVries:2018aul}, so we focus only on those from the pseudoscalar 
currents of the heavy flavors $f=t,\tau',\nu'$, when discussing the BAU in the SM4 below.

\begin{figure}[tbp]
\centering 
\includegraphics[scale=0.7]{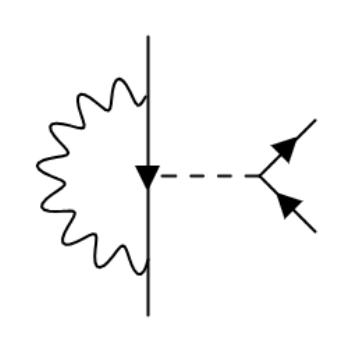}\hspace{1.0cm}
\includegraphics[scale=0.7]{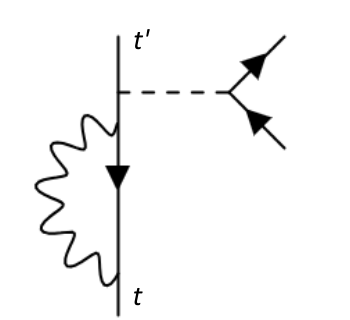}

(a)\hspace{5.0cm}(b)
\caption{\label{fig3}
(a) Vertex correction and (b) self-energy correction contained in the pseudoscalar penguin.}
\end{figure}

Evaluating a three-loop diagram directly is tedious. Our strategy is to build the 
subdiagrams first, such as the pseudoscalar penguin which involves the vertex correction in 
Fig.~\ref{fig3}(a) and the self-energy correction associated with the $t'\to t$ transition 
in Fig.~\ref{fig3}(b), and then insert it into the heavy quark loop to get the final 
result. The two subdiagrams are calculated in Appendix~\ref{appA}. We remind that the major 
source of theoretical uncertainties in the analysis resides in the formulation the transport 
equations, so this simplified handling should be acceptable. As illuminated in 
Appendix~\ref{appA}, both Figs.~\ref{fig3}(a) and \ref{fig3}(b) exhibit ultraviolet (UV) 
divergences, which turn out to cancel each other. Furthermore, their contributions cancel 
exactly in the limit $q^2\to 0$, $q^2$ being the invariant mass of the fermion pair 
$f\bar f$. It indicates that their net contribution is proportional to $q^2$, which then 
removes the pole of the virtual pseudoscalar propagator $1/q^2$ in the $R_\xi$ gauge with
the gauge parameter $\xi=0$, making a four-fermion operator 
$\bar t(1+\gamma_5)t'(\bar f \gamma_5f)$. Namely, 
the pseudoscalar penguin diagrams produce a four-fermion operator like the QCD and electroweak 
penguins in the effective weak Hamiltonian, whose appearances are warranted by the gauge 
invariance \cite{Silvestrini:2019sey}. 


The pseudoscalar penguin contribution derived in Appendix~\ref{appA} reads
\begin{eqnarray}
P=-\frac{i}{192}V_{t'b'}^*V_{tb'} I_{3b'}I_{3f}\frac{g_fg_{b'}^2g_{t'}}{m_{b'}^2}
(1+\gamma_5)\bar f(q_2)\gamma_5f(q_1),
\end{eqnarray}
where the external $t'$ and $t$ quark legs have been truncated, and $I_{3b'}$ ($I_{3f}$) is 
the third component of the $b'$ quark ($f$ fermion) isospin. Appendix~\ref{appA} also presents the 
expression for the self-energy correction associated with the $t\to t'$ transition, 
\begin{eqnarray}
S&=&\frac{i}{64\pi^2}V_{tb}^*V_{t'b} g_{t'}g_t\left(\ln\frac{\Lambda_s^2}{-p^2}+1\right)
\not p(1+\gamma_5),\label{25}
\end{eqnarray} 
with the $t$ quark momentum $p$. The electroweak symmetry restoration scale $\Lambda_s$
is introduced through the upper bound of the loop momentum, below which
the full theory with massive fourth generation quarks is applicable. 

We then come to the second step, computing the heavy quark loop with the insertions of 
the pseudoscalar penguin $P$ and the self-energy correction $S$. The loop 
integral with the $CP$-odd phase is written as 
\begin{eqnarray}
\Pi&=&-\frac{i}{64\pi^2}g_{t'}g_t
\frac{-i}{192}I_{3b'}I_{3f}\frac{g_fg_{b'}^2g_{t'}}{m_{b'}^2}
\int\frac{d^4l}{(2\pi)^4}\left(\ln\frac{\Lambda_s^2}{-l^2}+1\right)\nonumber\\
& &\times{\rm Tr}\left[i\not l(1+\gamma_5)
\frac{i(\not l+m_t)}{l^2-m_t^2}(1+\gamma_5)\frac{i(\not l+m_{t'})}{l^2-m_{t'}^2}
\frac{-ig_{t'}}{\sqrt{2}}\frac{i(\not l+m_{t'})}{l^2-m_{t'}^2}\right]J\bar f\gamma_5f,
\label{qp}
\end{eqnarray}
where the first minus sign on the right-hand side is attributed to the fermion loop, and Tr 
refers to the traces over the fermion and color flows. The external momentum from the Higgs 
boson and the electroweak-scale masses have been dropped relative to the heavy quark masses, 
except the $t$ quark mass $m_t$ in the numerator, which survives the trace. 
The imaginary part of the product of the $4\times 4$ CKM matrix elements,
\begin{eqnarray} 
J={\rm Im}(V^*_{t'b'}V_{tb'}V^*_{tb}V_{t'b}), \label{jar}
\end{eqnarray}
can be visualized as the area of the parallelogram formed by the vectors $V^*_{t'b'}V_{tb'}$ 
and $V^*_{tb}V_{t'b}$ in a complex plane. The trace gives a factor $4N_c=12$, $N_c$ being the 
number of colors. The integral in Eq.~(\ref{qp}) reduces to 
\begin{eqnarray}
\Pi&=&\frac{I_{3b'}I_{3f}g_fg_t^2m_{t'}^4}{32^2\pi^4\sqrt{2}v^4}
\left[1+Li_2\left(-\frac{\Lambda_s^2}{m_{t'}^2}\right)\right]J\bar f\gamma_5f,
\label{box}
\end{eqnarray}
where the loop momentum is cut at the restoration scale $\Lambda_s$, and the 
relations $m_{t,b',t'}=g_{t,b',t'}v/\sqrt{2}$ have been employed.

It has been known that the Yukawa couplings of fourth generation quarks evolve with the 
energy scale; they decrease with the scale as depicted in \cite{Hung:2009hy,Li:2025bon}.
The $t'$ quark mass $m_{t'}$ represents a characteristic scale of the heavy quark loop; 
for example, it characterizes the scale of the $t'\to t$ pseudoscalar
penguin in Fig.~\ref{fig3}(a). If $m_{t'}$ was above $\Lambda_s$, internal particles in 
the penguin diagram become massless, so the full theory with massive quarks does 
not apply. If $m_{t'}$ was much lower than $\Lambda_s$, such that $b'$ quarks form deep 
bound states, the formulation of the full theory is not appropriate either; fourth 
generation quarks are heavy enough for forming bound states at the electroweak scale as 
observed in \cite{Li:2023fim}. The above argument suggests to perform the matching between
the effective and full theories at the scale, where $m_{t'}$ takes a value right below 
$\Lambda_s$. Hence, we set $m_{t'}=\Lambda_s$, and have $Li_2(-1)=-\pi^2/12$. Equating the 
coefficient of the effective operator to Eq.~(\ref{box}) with $I_{3b'}=-1/2$, and defining 
the constant $c=1-\pi^2/12$, we arrive at
\begin{eqnarray}
\frac{s_fg_fv^2}{\sqrt{2}\Lambda^2}
=-\frac{c}{2}\frac{I_{3f} g_fg_t^2\Lambda_s^4J}{32^2\pi^4\sqrt{2}v^4}.\label{eq}
\end{eqnarray}
The value of the Jarlskog invariant $J$ will be estimated in the next section.

\section{$4\times 4$ CKM matrix}
\label{sec3}

We have conducted dispersive analyses on the mixing between the neutral quark states 
$Q\bar q$ and $\bar Qq$ \cite{Li:2023ncg}, where $Q$ ($q$) stands for a massive (light) 
quark. The case with $Q=c$ and $q=u$ corresponds to the $D$ meson mixing \cite{Li:2020xrz,Li:2022jxc}. 
The only assumption underlying the analyses is that the electroweak symmetry of the SM is 
restored as the heavy quark mass $m_Q$ is above a high energy scale \cite{Chien:2018ohd}. 
It has been demonstrated that this restoration can take place in our SM4 setup 
\cite{Li:2025bon}. The mixing phenomenon disappears owing to the unitarity 
of the CKM matrix, as the electroweak symmetry is restored. The disappearance of the 
mixing at $m_Q>\Lambda_s$, $\Lambda_s$ being the restoration scale, was then taken as 
the input to the dispersion relation for the mixing amplitude, whose behavior at low 
$m_Q<\Lambda_s$, i.e., in the symmetry broken phase, could be solved. It was found that 
the solution imposes stringent constrains on the quark masses and the CKM matrix elements 
appearing in the mixing amplitude \cite{Li:2023ncg}. The similar formalism works for the 
mixing between the lepton states $\mathcal{L}^-\ell^+$ and $\mathcal{L}^+\ell^-$
\cite{Li:2023ncg,Li:2024awx}, where $\mathcal{L}$ ($\ell$) denotes a massive (light) 
charged lepton. The case with $\mathcal{L}=\mu$ and $\ell=e$ corresponds to the mixing of 
muonia $\mu^-e^+$ and $\mu^+ e^-$. Some of our observations are that the normal ordering 
of the neutrino masses is favored, and that the smaller mixing angles in the quark sector 
than in the lepton sector are due to the hierarchical mass ratios 
$m_s^2/m_b^2\ll m_2^2/m_3^2$, where $m_s$, $m_b$, $m_2$ and $m_3$ are the masses of a 
strange quark, a bottom quark, a second-generation neutrino and a third-generation neutrino, 
respectively.

We extend the dispersive study on the $3\times 3$ CKM matrix in \cite{Li:2023ncg} to the 
one on the $4\times 4$ matrix. The lower mixing angles in the quark sector 
justify the Wolfenstein parametrization, i.e., the expansion of the CKM matrix elements in 
the Wolfenstein parameter $\lambda$. We follow the parametrization of the $4\times 4$ CKM 
matrix elements associated with fourth generation quarks in \cite{Alok:2010zj},
\begin{eqnarray}
V_{ub'}=p\lambda^3e^{-i\delta'},\;\;\;\;
V_{cb'}=q\lambda^2e^{-i\delta''},\;\;\;\;
V_{tb'}=r\lambda,\label{vub}
\end{eqnarray}
up to leading powers of $\lambda$. A $3\times 3$ ($4\times 4$) unitary 
matrix depends on three angles and one $CP$-odd phase (six angles and three $CP$-odd phases).  
We will treat the four parameters in the $3\times 3$ CKM matrix as inputs and solve for the 
remaining five unknowns $p$, $q$, $r$, $\delta'$ and $\delta''$. We have ensured 
that the alternative parametrization with $\delta''$ assigned to $V_{tb'}$
leads to the same Jalskog invariant.

We prepare the other matrix elements in powers of $\lambda$ to the desired accuracy,
quoting the results in \cite{Ahn:2011fg},
\begin{eqnarray}
V_{ud}=1-\frac{\lambda^2}{2}-\frac{\lambda^4}{8},\;\;\;
V_{us}=\lambda,\;\;\;\;
V_{ub}=A\lambda^3Ce^{-i\delta},\label{vud}
\end{eqnarray}
with 
\begin{eqnarray}
C=\sqrt{\rho^2+\eta^2}\approx \frac{\sqrt{\bar\rho^2+\bar\eta^2}}{1-\lambda^2/2}
\end{eqnarray} 
in terms of the conventional parameters $\rho$ and $\eta$,
and the weak phase $\delta$ in the $3\times 3$ CKM matrix. Equation~(\ref{vud})
together with $V_{ub'}$ in Eq.~(\ref{vub}) obey the unitarity condition for the first 
row of the matrix up to corrections of $O(\lambda^6)$. The impact on $V_{ud}$ from the 
additional $V_{ub'}$ starts at $O(\lambda^6)$,
\begin{eqnarray}
V_{ud}=1-\frac{\lambda^2}{2}-\frac{\lambda^4}{8}
-\frac{\lambda^6}{16}(1+8A^2C^2+8p^2),
\end{eqnarray}
which improves the unitarity condition up to corrections of $O(\lambda^7)$.

The elements of the second row are written as
\begin{eqnarray}
V_{cd}&=&-\lambda+\frac{\lambda^5}{2}
(A^2+q^2-2A^2Ce^{i\delta}-2pqe^{i\delta'-i\delta''}),\nonumber\\
V_{cs}&=&1-\frac{\lambda^2}{2}-\frac{\lambda^4}{8}(1+4A^2+4q^2),\;\;\;\;
V_{cb}=A\lambda^2.\label{vcd}
\end{eqnarray}
It is trivial to verify the unitarity conditions, i.e., the normalization of the 
second row and its orthogonality to the first row, up to $O(\lambda^5)$.
The elements of the third row are given by
\begin{eqnarray}
V_{td}&=&A\lambda^3(1-Ce^{i\delta})+r
(qe^{i\delta''}-pe^{i\delta'})\lambda^4+\frac{A}{2}[(1+r^2)Ce^{i\delta}-r^2]\lambda^5,\nonumber\\
V_{ts}&=&-A\lambda^2-qre^{i\delta''}\lambda^3+\frac{A}{2}
(1+r^2-2Ce^{i\delta})\lambda^4+\frac{1}{2}r
(qe^{i\delta''}-2pe^{i\delta'})\lambda^5,\nonumber\\
V_{tb}&=&1-\frac{r^2}{2}\lambda^2-\frac{1}{8}(4A^2+r^4)\lambda^4
-Aqr\cos\delta''\lambda^5.\label{vtd}
\end{eqnarray}
The explicit expressions for the fourth row elements $V_{t'i}$, $i=d,s,b,b'$, can be acquired 
via the unitarity conditions.
 
The formulas for the lepton state mixing, such as the $\mu^- e^+$-$\mu^+ e^-$ muonium mixing, 
in the SM4 \cite{Li:2024awx} are gathered in Appendix~\ref{appB}, where they are reinvestigated 
incidentally. The dispersive constraints from the quark state mixing on the quark masses and 
mixing angles are similar, as long as the electroweak symmetry is restored above the scale 
$\Lambda_s$; one simply replaces the neutrino masses $m_1$, $m_2$, $m_3$ and $m_4$ in 
Eqs.~(\ref{cs1})-(\ref{cs3}) and Eq.~(\ref{fs}) by the down-type quark masses $m_d$, $m_s$, 
$m_b$ and $m_{b'}$, respectively. It is obvious that the dispersive constraints can be 
satisfied by diminishing the common factors in the above equations,
\begin{eqnarray}
& &\lambda_d\frac{m_{b'}^2-m_d^2}{m_W^2-m_d^2}+\lambda_s\frac{m_{b'}^2-m_s^2}{m_W^2-m_s^2}
+\lambda_b\frac{m_{b'}^2-m_b^2}{m_W^2-m_b^2}\approx 0,\label{au0}\\
& &\frac{\lambda_d}{m_W^2-m_d^2}+\frac{\lambda_s}{m_W^2-m_s^2}
+\frac{\lambda_b}{m_W^2-m_b^2}\approx 0,\label{au}
\end{eqnarray}
with the product of the CKM matrix elements $\lambda_i=V^*_{Qi}V_{qi}$, $i=d$, $s$, $b$,
and the $W$ boson mass $m_W$. 

We implement the approximation
\begin{eqnarray}
\frac{m_{b'}^2-m_i^2}{m_W^2-m_i^2}\approx 
\frac{m_{b'}^2}{m_W^2}\left(1-\frac{m_i^2}{m_{b'}^2}+\frac{m_i^2} {m_W^2}\right)
\approx \frac{m_{b'}^2}{m_W^2}\left(1+\frac{m_i^2} {m_W^2}\right)
\approx \frac{m_{b'}^2}{m_W^2-m_i^2},
\end{eqnarray}
based on the mass hierarchy $m_i^2\ll m_W^2\ll m_{b'}^2$. Equations~(\ref{au0}) and 
(\ref{au}) then turn into the same form 
\begin{eqnarray}
\lambda_{b'}=\lambda_d\frac{m_d^2}{m_W^2}+\lambda_s\frac{m_s^2}{m_W^2}
+\lambda_b\frac{m_b^2}{m_W^2}\approx \lambda_s\frac{m_s^2}{m_W^2}
+\lambda_b\frac{m_b^2}{m_W^2},\label{ex1}
\end{eqnarray}
where the unitarity condition $\lambda_d+\lambda_s+\lambda_s=-\lambda_{b'}$ has been inserted,
and the $d$ quark mass $m_d$ has been ignored. Note that Eqs.~(\ref{au0}) and (\ref{au}) 
stand for two distinct constraints on the lepton mixing, since the masses of a fourth 
generation neutrino and a $W$ boson are of the same order, and the $m_i^2/m_4^2$ pieces, 
$i=1$, 2, 3, are comparable with $m_i^2/m_W^2$. Besides, the masses of down-type quarks 
(neutrinos) in the intermediate channels of the considered neutral fermion state mixings 
possess different spectra. The above explain why the dispersion relations produce variant 
solutions for the CKM and PMNS matrices.

Equation~(\ref{ex1}) will be utilized to determine the $4\times 4$ CKM matrix 
elements. The light quark in the proposed heavy-light system is regarded as being 
massless \cite{Li:2023ncg}. We do not include the $t'\bar u$-$\bar t' u$ mixing; the mass 
$m_{t'}\approx 200$ TeV of a superheavy $t'$ quark at the electroweak scale, much higher 
than the symmetry restoration scale $\Lambda_s$ definitely \cite{Li:2025bon}, renders all 
internal particles in the box diagrams massless, so the mixing phenomenon disappears. In 
other words, the $t'\bar u$-$\bar t' u$ mixing is not a suitable process for the 
study. The real and imaginary parts of the dispersion relations in Eq.~(\ref{ex1}) 
from the $c\bar u$-$\bar c u$ and $t\bar u$-$\bar t u$ mixings impose four constraints. 
To have enough equations for the five unknowns $p$, $q$, $r$, $\delta'$ and 
$\delta''$, we must take into account the $t\bar c$-$\bar t c$ mixing. A $tc$ pair is not a 
heavy-light system, strictly speaking, to which the present formulas may not hold well. In 
principle, we should construct a more complicated mixing amplitude for the $tc$ case, in 
which both external quarks have finite masses. We will not attempt such a rigorous approach 
here, but treat a charm quark as being massless. The theoretical uncertainty stemming from 
this approximation is anticipated to be minor relative to those in the formulation for 
the EWBG.

The $c\bar u$-$\bar c u$, $t\bar u$-$\bar t u$ and $t\bar c$-$\bar t c$ mixing amplitudes, 
each of which involves real and imaginary parts, offer six relations. That is, our setup 
is overconstrained. We first exemplify that when the strange quark mass $m_s$, which can be 
regarded as the sixth unknown, vanishes, a unique set of solutions for $p$, $q$, $r$, 
$\delta'$ and $\delta''$ can be established. These solutions give an idea on the order of 
magnitude of the matrix elements associated with a fourth generation. Employing 
Eqs.~(\ref{vub}), (\ref{vud}), (\ref{vcd}) and (\ref{vtd}), we attain, from Eq.~(\ref{ex1}),
\begin{eqnarray}
& &pqe^{-i(\delta'-\delta'')}=\frac{m_b^2}{m_W^2}A^2Ce^{-i\delta},\nonumber\\
& &pre^{-i\delta'}=\frac{m_b^2}{m_W^2}\frac{AC}{\lambda}
\left(1-\frac{r^2\lambda^2}{2}\right)e^{-i\delta},\nonumber\\
& &qre^{i\delta''}=\frac{m_b^2}{m_W^2}\frac{A}{\lambda}
\left(1-\frac{r^2\lambda^2}{2}\right),\label{con2}
\end{eqnarray}
for the $c\bar u$-$\bar c u$, $t\bar u$-$\bar t u$ and $t\bar c$-$\bar t c$ mixings,
respectively. 

It is straightforward to solve for
\begin{eqnarray}
p=\frac{m_bAC}{m_W},\;\;q=\frac{m_bA}{m_W},\;\;r=\frac{m_b}{\lambda m_W},\;\;
\delta'=\delta,\;\;\delta''=0.
\end{eqnarray}
The values of $p$, $q$ and $r$ are all of order of $m_b/m_W$ with the inequality
$p<q<r$. The second terms $r^2\lambda^2/2$ in the parentheses of Eq.~(\ref{con2}), being 
suppressed by the power $m_b^2/m_W^2$ compared to the first terms, are negligible. We have
the four inputs \cite{PDG} 
\begin{eqnarray}
\lambda\approx 0.225,\;\; A\approx 0.826,\;\;C\approx 0.397,\;\;\delta\approx 1.15,
\end{eqnarray}
among which the value of $\lambda$ is precise, $A$ and $\delta$ have about 2\% uncertainties, 
and $C$ can vary by 3\%. The $W$ boson mass is set to $m_W\approx 80.4$ GeV \cite{PDG}, 
which has much higher precision. Because our analysis is based on the tree-level expression 
for the box diagrams, it makes sense to choose a scale-independent quark mass.
We adopt the pole mass $m_b\approx 4.78$ GeV for a $b$ quark with roughly 1\% 
uncertainty \cite{PDG}, deriving the matrix elements in Eq.~(\ref{vub})
\begin{eqnarray}
V_{ub'}\approx 2.22\times 10^{-4}e^{-1.15 i},\;\;
V_{cb'}\approx 2.49\times 10^{-3},\;\;
V_{tb'}\approx 5.95\times 10^{-2}.
\end{eqnarray}
The results indicate that a more off-diagonal matrix element has a smaller magnitude
in consistency with the pattern of the $3\times 3$ matrix.

Next we take the finite mass $m_s$, for which one of the six constraints needs to be 
relaxed. Viewing the smallness of the phase $\delta''$ and the potential uncertainty from 
ignoring the charm quark mass, we exclude the imaginary part of the constraint from the 
$t\bar c$-$\bar t c$ mixing. Equation~(\ref{con2}) becomes
\begin{eqnarray}
& &pqe^{-i(\delta'-\delta'')}=\frac{m_b^2}{m_W^2}A^2Ce^{-i\delta}
+\frac{m_s^2}{m_W^2}\frac{1}{\lambda^4}\,\nonumber\\
& &pre^{-i\delta'}=\frac{m_b^2}{m_W^2}\frac{AC}{\lambda}
e^{-i\delta}-\frac{m_s^2}{m_W^2}\frac{A}{\lambda},\nonumber\\
& &qr\cos\delta''=\frac{m_b^2}{m_W^2}\frac{A}{\lambda}
-\frac{m_s^2}{m_W^2}\frac{A}{\lambda},
\end{eqnarray}
where the $r^2\lambda^2/2$ terms have been dropped, and only the $m_s$-dependent 
terms leading in powers of $\lambda$ are retained. It is seen that the $m_s$-dependent correction 
mainly occurs in the first relation, which is associated with the large product of the
matrix elements $V^*_{us}V_{cs}$. The $m_s$-dependent terms in the second and third 
lines have tiny effects. 


It is tricky to choose a scale-independent strange quark mass. A pole mass is not well defined 
for a light quark owing to the unreliable perturbative renormalzation-group (RG) 
evolution at a low scale. We assume that the running coupling constant is frozen at a low 
scale by, for instance, a gluon mass induced by confinement. In this way the strange quark
$\overline{\rm MS}$ running mass is also frozen, and serves as a scale-independent mass. 
It is thus reasonable to set the strange quark mass to its $\overline{\rm MS}$ value at the 
renormalization scale $\mu=1$ GeV \cite{PDG}, $m_s=0.11\pm 0.02$ GeV with 20\% uncertainty, 
which covers the variation within the range 0.5 GeV$^2<\mu^2<4.0$ GeV$^2$. The coupled 
equations are solved numerically to yield
\begin{eqnarray}
& &p=0.0223^{+0.0001}_{-0.0004},\;\; q=0.0639_{+0.0085}^{-0.0053},\;\; 
r=0.231^{+0.015}_{-0.020},\nonumber\\
& &\delta'=1.15\pm 0.03,\;\; \delta''=0.489_{+0.122}^{-0.137}, 
\end{eqnarray}
whose uncertainties originate from those of the Wolfenstein parameters and of the $b$ and $s$ 
quark masses. The $s$ quark mass is responsible for the dominant source of uncertainties. 

The fourth column elements are then given by
\begin{eqnarray}
V_{ub'}&=&2.54^{+0.02}_{-0.05}\times 10^{-4}\exp[-(1.15\pm 0.03) i],\nonumber\\
V_{cb'}&=&3.23_{+0.43}^{-0.26}\times 10^{-3}\exp[-(0.489_{+0.122}^{-0.137}) i],\nonumber\\
V_{tb'}&=&5.20^{+0.34}_{-0.45}\times 10^{-2}.
\label{res}
\end{eqnarray}
It is noticed that the inclusion of the $s$ quark mass modifies the phase 
$\delta''$ mostly, and has negligible impacts on the magnitude of the matrix elements. 
The evaluation of the Jarlskog invariant in Eq.~(\ref{jar}) demands the input of the 
element $V_{t'b}$. For completeness, we collect our predictions for the fourth row of the 
$4\times 4$ CKM matrix based on the unitarity,
\begin{eqnarray}
V_{t'u}&=&3.16^{-1.05}_{+1.32}\times 10^{-4}\exp[(1.15^{+0.03}_{-0.03})i],\nonumber\\
V_{t's}&=&-(1.63^{-0.55}_{+0.69})\times 10^{-3}\exp[(1.15^{+0.03}_{-0.03})i],
\nonumber\\
V_{t'b}&=&-(5.20^{+0.34}_{-0.44})\times 10^{-2}
\exp[(1.22^{-0.44}_{+0.62})\times 10^{-3}i],\nonumber\\
V_{t'b'}&=&0.999_{+0.000}^{-0.001}.
\label{re4}
\end{eqnarray}
The above solutions evince the maximal unitarity violation in the third row of the $3\times 3$ 
CKM matrix, and the potential of searching for the unitarity violation in $B$ meson decays.

We confront our predictions for the fourth column and row of the $4\times 4$ CKM matrix with
those extracted from measurements in the kaon and $B$ meson systems \cite{Kaur:2024jlo},
\begin{eqnarray}
& &|V_{ub'}|=0.017\pm 0.014,\;\;
|V_{cb'}|=(8.4\pm 6.2)\times 10^{-3},\;\;
|V_{tb'}|=0.07\pm 0.08,\nonumber\\
& &|V_{t'd}|=0.01\pm 0.01,\;\;|V_{t's}|=0.01\pm 0.01,\;\;|V_{t'b}|=0.07\pm 0.08,\nonumber\\
& &|V_{t'b'}|=0.998\pm 0.006.
\end{eqnarray}
It is apparent that the experimental bounds are still loose and broad ranges of
the matrix elements are allowed. The bounds on the $CP$-odd phases are even weaker
\cite{Kaur:2024jlo},
\begin{eqnarray}
\delta'=1.21\pm 1.59,\;\;\delta''=1.10\pm 1.64.
\end{eqnarray} 
The results in Eqs.~(\ref{res}) and (\ref{re4}) comply with the current experimental 
constraints.

The $2.3\sigma$ deviation from the unitarity in the first row of the $3\times 3$ CKM matrix
\begin{eqnarray}
|V_{ud}|^2+|V_{us}|^2+|V_{ub}|^2=0.9984\pm 0.0007,
\end{eqnarray}
was reported \cite{PDG}, known as the Cabibbo angle anomaly. The unitarity of the second row, 
the first column and the second column of the $3\times 3$ CKM matrix have been scrutinized to the 
precision of $O(10^{-4})-O(10^{-2})$ \cite{Kitahara:2024azt,Gorchtein:2025wli} as summarized 
in \cite{PDG},
\begin{eqnarray}
& &|V_{cd}|^2+|V_{cs}|^2+|V_{cb}|^2=1.001\pm 0.012,\nonumber\\
& &|V_{ud}|^2+|V_{cd}|^2+|V_{td}|^2=0.9971\pm 0.0020,\nonumber\\
& &|V_{us}|^2+|V_{cs}|^2+|V_{ts}|^2=1.003\pm 0.012.
\end{eqnarray}
Equations~(\ref{res}) and (\ref{re4}) hint what experimental precision need to be 
reached in order to confirm the deviations from unity \cite{Seng:2021gmh}.

\section{Baryon asymmetry}
\label{sec4}

We now predict the Jarlskog invariant for the CPV source in the dimension-6 operators,
\begin{eqnarray}
J=-(3.30^{-0.89}_{+0.85})\times 10^{-6},\label{jal}
\end{eqnarray}
from the $4\times 4$ CKM matrix elements obtained in the previous section.
It has been checked that the pairs of matrix element products $V_{cb}^*V_{cb'}$ and
$V_{t'b}^*V_{t'b'}$ turn in the same Jarlskog invariant 
${\rm Im}(V_{cb}^*V_{cb'}V_{t'b}V_{t'b'}^*)=+3.30\times 10^{-6}$ but with an opposite sign. 
It should be the case, for the shortest side $V_{ub}^*V_{ub'}$ of the parallelogram has 
negligible length. We have also inspected the areas defined by the other pairs of matrix 
element products associated with fourth generation quarks, and affirmed that the one in 
Eq.~(\ref{jal}) is the maximum. As already stated, substituting a $c$ ($s$) quark for 
the $t$ ($b$) quark in Fig.~\ref{fig2}(b) causes suppression by the $c$ ($s$) quark Yukawa 
coupling. Hence, the transition $t'\to b'\to t\to b\to t'$ described in Fig.~\ref{fig2}(b) 
indeed captures the leading CPV source. 

Note that the Jarlskog invariant for the $4\times 4$ CKM matrix in Eq.~(\ref{jal}) 
is negative. To guarantee a positive $\Lambda^2$ in Eq.~(\ref{eq}), we must relate the 
sign $s_f$ to the third component $I_{3f}$ of the fermion $f$, $s_f=2I_{3f}$ ($I_{3f}$ 
has the magnitude of $1/2$). The relation accords with the choice $s_t=+1$ ($s_\tau=-1$) 
for the top-sourced ($\tau$-sourced) operator in \cite{DeVries:2018aul} inferred by
the observed BAU. It is worth emphasizing that the sign $s_f$ is not a fitting 
parameter, but can be specified in our formalism. Equation~(\ref{eq}) then sets 
the new physics scale
\begin{eqnarray}
\Lambda
=\frac{64\pi^2 v^3}{g_t\sqrt{-cJ}\Lambda_s^2},\label{lam}
\end{eqnarray}
manifesting that $\Lambda$ is independent of the flavor $f$, contrary to the treatment
in \cite{DeVries:2018aul,Huber:2006ri}. 


Equation~(\ref{lam}) requires the knowledge of the symmetry restoration scale $\Lambda_s$ for 
estimating the new physics scale $\Lambda$, which can be determined in a RG analysis on the 
relevant Yukawa couplings. Though the evolutions of the Yukawa couplings have been probed 
in \cite{Li:2025bon}, the exact scale for the symmetry restoration remains unaddressed 
in the SM4. It was shown \cite{Li:2023fim} that fourth generation quarks with order-of-TeV 
masses form bound states at a low scale, and that the heavy bound states couple to 
Higgs bosons more weakly. It means that heavy 
quark effects ought to be excluded, when the RG equations are solved 
around the electroweak scale. Heavy quarks, as physical degrees of freedom at a high 
scale, contribute to the RG evolution above the restoration scale. Such a two-stage RG 
study was not done in \cite{Li:2025bon}; the purpose of \cite{Li:2025bon} is  
to explore the Yukawa couplings of fourth generation quarks around the UV fixed point, 
and to certify that they are below the critical values for establishing condensates, and 
the electroweak symmetric phase exists at a high energy in the SM4.

We pinpoint the restoration scale $\Lambda_s$ in the SM4 by solving the RG equations for 
the Yukawa couplings in two regions of the evolution variable $x\equiv\ln(\mu/m_Z)$ 
\cite{Hung:2009hy}, $\mu$ being a renormalization scale and $m_Z$ being the $Z$ boson mass. 
In the first region $x$ runs from $x=0$, corresponding to the electroweak scale, i.e., the 
starting point of the RG evolution, to some value $x=x_s$. We switch off the heavy quark 
contributions in this region with $0\le x<x_s$. Note that the degeneracy of the fourth 
generation masses was assumed in the derivation of the RG equations in \cite{Hung:2009hy}. 
We set the initial condition of the heavy lepton Yukawa coupling $g_L$ at $x=0$ to 
$g_{\tau'}=\sqrt{2} m_{\tau'}/v=1.6$ of $\tau'$ leptons for illustration. We have assured 
that the above simplification does not affect the outcome much. The value $g_t=1.0$ is 
assigned to the initial condition of the $t$ quark Yukawa coupling at $x=0$. In the second 
region $x$ runs from $x=x_s$ to, say, $x=12$, which has reached the UV fixed point. The 
heavy quark Yukawa coupling $g_Q$ is taken into account in this region, and chosen as the 
critical value $g_Q^c=9.1$ at $x=x_s$ \cite{Li:2025bon}. We then adjust $x_s$, such that 
the Yukawa couplings $g_L$ and $g_t$ from the two regions match at $x=x_s$ smoothly. Once 
$x_s$ is identified, we read off the restoration scale $\Lambda_s=m_Z\exp(z_s)$.

\begin{figure}[tbp]
\centering 
\includegraphics[scale=0.4]{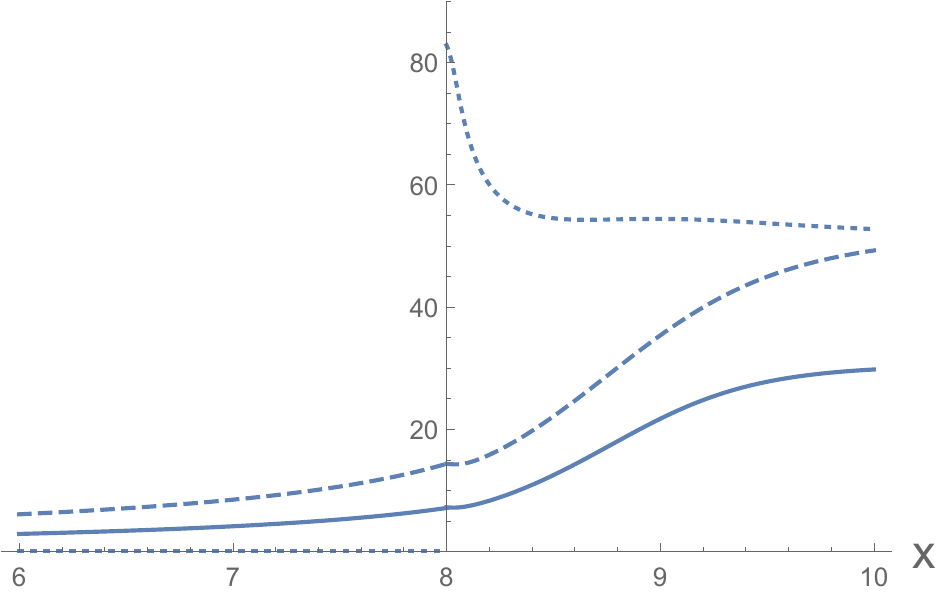}\hspace{1.0cm}
\includegraphics[scale=0.4]{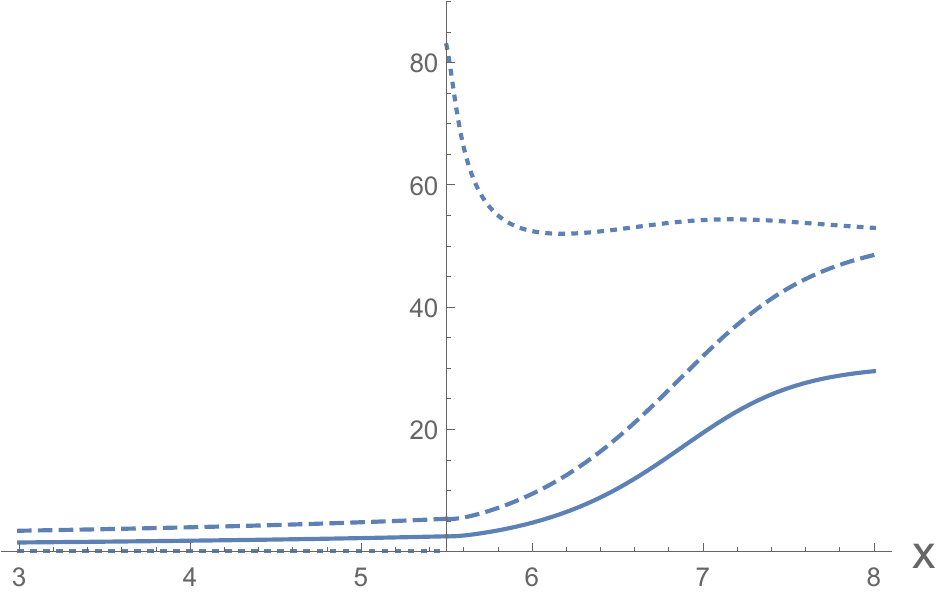}

(a)\hspace{7.0cm}(b)

\caption{\label{fig4}
RG evolutions of the squares of the Yukawa couplings $g_Q^2$ (dotted),
$g_L^2$ (dashed) and $g_t^2$ (solid) matched at (a) $x_s=8.0$ and (b) $x_s=5.5$. }
\end{figure}


We do not involve the running of the gauge couplings here for simplicity, which are 
smaller than the considered Yukawa couplings.
The matchings of the squares of Yukawa couplings $g_L^2$ and $g_t^2$ at $x=x_s=8.0$ are 
displayed in Fig.~\ref{fig4}(a) to reveal the jagged transition of the RG evolutions from 
$x<x_s$ to $x>x_s$. Looking at the behaviors around $x=8.0$ closely, one finds that the 
curves go down and up in the interval $8.0<x<8.3$. The matching becomes smooth as $x_s$ 
decreases to a value between $x_s=6.0$ and $x_s=5.0$, so the choice $x_s=5.5$ is reasonable, 
for which the RG evolutions of $g_Q^2$, $g_L^2$ and $g_t^2$ are shown in Fig.~\ref{fig4}(b). 
It is seen that $g_L^2$ and $g_t^2$ ascend with $x$ monotonically in the first region. In 
the second region $g_Q^2$ runs from its critical value $g_Q^{c2}=83$ \cite{Li:2025bon} down 
to the fixed-point value, approaching $g_Q^{f2}= 52$ at a high $x$
\cite{Hung:2009hy}, the same as observed in \cite{Li:2025bon}. The curves of 
$g_L^2$ and $g_t^2$ in the two regions meet smoothly at $x=5.5$ with the magnitudes 
$g_L^2=5.4$ and $g_t^2=2.5$, respectively. The significant enhancement of the slopes at 
$x>x_s$ originates from the inclusion of the heavy quark effects. The curve of $g_L^2$
merges with that of $g_Q^2$, and also approaches the fixed-point value $g_L^{f2}= 52$
at a high $x$. The top quark Yukawa coupling squared takes the fixed-point value
$g_t^{f2}=30$. Figure~\ref{fig4}(b) designates the restoration scale 
$\Lambda_s=m_Z\exp(z_s)\approx 22$ TeV for $m_Z=91.2$ GeV \cite{PDG}, in agreement with the 
order-of-magnitude estimate in the dispersive analysis on the lepton mixing \cite{Li:2024awx}.

As elaborated before, the effective operators are matched to the full theory near the 
restoration scale $\Lambda_s$. We then deduce, from Eq.~(\ref{lam}) for $g_t=\sqrt{2.5}$ 
at $x=x_s=5.5$ and $v=246$ GeV,
\begin{eqnarray}
\Lambda\approx 16\;{\rm TeV},\label{16t}
\end{eqnarray}
which is above the lower bounds $\Lambda_t=7.1$ ($\Lambda_\tau=1.0$) TeV for $t$ quarks 
($\tau$ leptons) from the electron EDM data \cite{DeVries:2018aul}. 
An upper bound $\Lambda_\tau< 3.6$ TeV was concluded in \cite{Fuchs:2020uoc} for the 
$\tau$ lepton operator to accommodate the observed BAU, which is below 
Eq.~(\ref{16t}). A more stringent constraint from the electron EDM based on the assumption 
of Minimal Flavor Violation, $\Lambda_\tau>6$ TeV, was imposed \cite{Alonso-Gonzalez:2021jsa}. 
Anyway, it is quite certain that the BAU sourced by $\tau$ leptons is insufficient in 
the SM4. Therefore, we turn to the BAU sourced by fourth generation leptons. 

When the temperature drops to the critical temperature, the Yukawa interaction 
among fourth generation quarks becomes strong enough, such that fourth generation quarks 
begin to form condensates in some regions of space \cite{Li:2025bon}. In these regions 
the electroweak symmetry is broken, a nontrivial vacuum with a finite VEV for a Higgs 
field is developed, particles acquire masses proportional to the VEV, and heavy scalars 
emerge as bound states of massive fourth generation quarks, whose interaction with Higgs 
bosons triggers the first-order EWPT. A bubble wall is then built up between the regions 
with the trivial vacuum (symmetric phase) and with the nontrivial vacuum (broken phase). 
Following bubble nucleation, the potential energy differential  
drives the bubble wall to accelerate into the symmetric phase. A massless fourth 
generation quark in the symmetric phase collides with the wall, as the bubble expands. When 
it crosses the wall, its mass increases with the space-dependent VEV, varying from the trivial 
vacuum to the nontrivial one. Owing to its large Yukawa coupling, the increase of the mass is 
so rapid that its momentum cannot sustain the large mass under the energy conservation soon 
after it penetrates into the wall. Note that the maximal mass it can get is the magnitude 
of its momentum  \cite{DeCurtis:2022hlx}. Therefore, most of fourth generation quarks are 
reflected back to the symmetric phase, exerting a frictional force on the wall, which slows 
down the bubble expansion. The slowdown then permits more time for the energy dissipation, so
the wall width increases \cite{Branchina:2025jou}.

Lighter particles, when colliding with a bubble wall, are transmitted 
partially into the broken phase and reflected partially back to the symmetric phase. They 
also exert a frictional force on a wall. Note that the potential difference between the two 
sides of a wall is related to particle degrees of freedom in each phase \cite{Branchina:2025jou}.
Since there are just two more fourth generation quarks on top of six SM quarks, the former 
are not expected to modify the wall velocity dramatically; the effect is perhaps at the 
30\% level, and should not be crucial for our estimate on the BAU, compared to other
sources of theoretical uncertainties in the formalism. A quantitative study of the impact 
on the wall expansion caused by heavy quark collisions goes beyond the scope of the present 
work. For a brief introduction to how to calculate a wall velocity in the local thermal 
equilibrium approximation, refer to \cite{Carena:2025flp}. Its application to probing the
wall dynamics in the SM4 setup can be pursued in the future.

According to Eq.~(14) in \cite{DeVries:2018aul}, we can construct the transport equations for 
the net number densities of left-handed fourth generation leptons, 
$L=n_{\nu'_{L}}+n_{\tau'_L}$, and of right-handed fourth generation leptons, $n_{\nu'_{R}}$ 
and $n_{\tau'_R}$. A net number density is defined as the difference between particle and 
antiparticle number densities. The $\nu'$ mass $m_4\approx 170$ GeV is as large as the top 
quark mass $m_t$, so both the $\nu'_L$ and $\nu'_R$ net number densities are relevant. 
The equations are similar to those for the third generation leptons, 
$l=n_{\nu_L}+ n_{\tau_L}$, $n_{\nu_R}$ and $n_{\tau_R}$, respectively. We put aside the 
potential quark-lepton interactions as done in Sections~3 and 4 of \cite{DeVries:2018aul}. 
The source terms appearing in the transport equations sensitively depend on the 
methods for their derivations \cite{Cline:2021dkf}, such as the VEV-insertion 
approximation (VIA) \cite{Lee:2004we} and the WKB expansion 
\cite{Joyce:1994fu,Cline:2000nw,Cline:2017jvp}; the VIA could lead to a baryon asymmetry 
orders of magnitude greater than the WKB formalism. Namely, tremendous theoretical 
uncertainties in predictions for the BAU are unavoidable, and an adequate estimation serves 
the purpose. It has been surveyed \cite{DeVries:2018aul} that the predicted 
baryon asymmetries in the above two approaches differ by a factor of 0.2-5, varying with 
the flavors responsible for the source terms. The VIA method was implemented in 
\cite{DeVries:2018aul}.

The coupled transport equations for various net number densities of quarks and leptons have 
been solved in \cite{DeVries:2018aul} with a plane bubble wall located at $z=0$, the 
symmetric phase in the $z<0$ region, and the broken phase in the $z>0$ region. 
The distinction between the transport equations for lepton and quark number densities 
mainly arises from strong sphaleron interactions. Strong sphalerons tend to wash out the 
chiral asymmetry in quark, instead of lepton, number densities generated during plasma 
particle collisions with bubble walls and diffusing into the symmetric phase 
\cite{Mohapatra:1991bz,Giudice:1993bb}. As depicted in Fig.~2 of \cite{DeVries:2018aul}, the 
net number density $l$ of left-handed third generation leptons plateaus at a large distance 
from a bubble wall due to the absence of strong sphaleron effects. It stays constant roughly 
at $-z>0$ till a distance of order of the diffusion length. It is expected that the solutions 
for the $\tau'_L$ and $\nu'_L$ net number densities also plateau at finite $-z$, and start to 
descend at a distance characterized by the lepton diffusion length. That is, both the 
$\tau'_L$ and $\nu'_L$ net number densities behave like $l$, whose analytical expression has 
been approximated by Eq.~(26) in \cite{DeVries:2018aul}. The net number density $l$ 
is negative, a consequence directly related to the sign of the CPV source. As a contrast, 
quark net number densities drop quickly with $-z$ in Fig.~2 of \cite{DeVries:2018aul}, 
signifying shorter quark diffusion lengths. A quark net number density would exhibit a feature 
similar to that of a lepton net number density, if there were no strong sphalerons. 

Weak sphaleron effects with slow rates are activated after the transport equations are 
solved according to the two-step process \cite{DeVries:2018aul}. In principle, the reduction 
of left-handed quark net number densities by weak sphalerons makes an excess of 
right-handed quark net number densities in front of bubble walls, which is then absorbed by 
expanding bubbles, giving rise to the baryon asymmetry in the broken phase. The chiral 
asymmetry (or equivalently, $CP$ asymmetries) in net number densities is turned into the 
baryon asymmetry through this mechanism. Nevertheless, the quark sourced chiral asymmetry 
would be fast redistributed between left- and right-handed quarks by strong sphalerons as 
reiterated before, and not efficient. Lepton chiral asymmetries are barely redistributed 
by strong sphalerons, and not washed out. It has been stressed that the left-handed 
lepton net number densities are negative. Since weak sphalerons preserve the number 
difference $B-L$ between left-handed baryons and leptons, the reduction of left-handed 
lepton net number densities creates an excess of quark net number densities. For 
a formulation of the baryon asymmetry as a result of weak sphaleron effects on lepton net 
number densities, refer to Appendix B of \cite{DeVries:2018aul}. More details can 
be looked up in \cite{Fuchs:2020pun,Joyce:1994zn}. 

It has been admitted that the CPV from the top quark source cannot account for the observed 
BAU \cite{DeVries:2018aul}. The numerical analysis in \cite{DeVries:2018aul} for a set of 
benchmark parameters discloses that top quarks contribute only about 1\% of the 
baryon-over-entropy ratio $\eta_B$, i.e., $\eta_B^{(t)}\approx 1.6\times 10^{-12}$ (see 
Table~3 and Fig.~2 in \cite{DeVries:2018aul}). The contributions from lighter quarks, such as 
$b$ quarks, are even lesser. Light neutrino contributions are also negligible, and light 
right-handed neutrinos decouple because of their tiny masses and Yukawa couplings. On the 
contrary, $\tau$ leptons, which have a finite Yukawa coupling and do not suffer the strong 
sphaleron suppression, play a more crucial role \cite{Chung:2009cb}; they induce the BAU 
$\eta_B= 8.2\times 10^{-11}$ at the correct order of magnitude for the new physics scale 
$\Lambda_\tau= 1$ TeV and the same set of benchmark parameters in \cite{DeVries:2018aul}. 
We have pointed out that the scale $\Lambda_\tau= 1$ TeV solely from the experimental 
constraint may be underestimated, and that the new physics scale associated with $\tau$ 
leptons in the SM4 is much higher.

We postulate that fourth generation leptons in the SM4, substituting for $\tau$ leptons 
\cite{DeVries:2018aul}, produce the observed BAU. We need not solve the coupled transport 
equations for the SM4, but extract the predictions from the framework developed 
by \cite{DeVries:2018aul} based on the effective theory. Given the analytical solution in 
Eq.~(26) of \cite{DeVries:2018aul}, the baryon-over-entropy ratio in Eq.~(28) of 
\cite{DeVries:2018aul} scales with the strength of the CPV source,
\begin{eqnarray}
\eta_B\propto\frac{y_f}{\Lambda^2},\label{str}
\end{eqnarray} 
as indicated in Eq.~(33) of \cite{DeVries:2018aul}. The benchmark parameters listed in Table~2 
of \cite{DeVries:2018aul} justifies the above scaling law. We take the same set of benchmark 
parameters as in \cite{DeVries:2018aul}, assuming the roughly equal relaxation rates, Yukawa 
rates, i.e., equal diffusion length and interaction length for $\tau'_L$ and $\nu'_L$, 
such that their contributions to the BAU are described by the scaling law in Eq.~(\ref{str}).

The entries in Table~1 of \cite{DeVries:2018aul}, i.e., the predicted 
$\eta_B=(7.3$-$8.3)\times 10^{-11}$, come from the numerical solutions to the transport 
equations with the $\tau$ lepton source for the Yukawa coupling $g_\tau=0.01$ and the 
new physics scale $\Lambda_\tau=1.0$ TeV. Both left-handed leptons $\tau'_L$ and $\nu'_L$ 
contribute to the BAU in our setup. The ratio of the CPV strengths 
\begin{eqnarray}
\frac{(g_{\tau'}+g_{\nu'})/\Lambda^2}{g_{\tau}/\Lambda_{\tau}^2}\approx 1,
\end{eqnarray}
then implies the similar outcome $\eta_B\approx 10^{-10}$ in the SM4.
We do not intend to present a precise number, viewing the potential large
uncertainties from, for instance, the estimate of the CPV sources, the determination
of the symmetry restoration scale, the variation of involved length scales, the calculations 
of the bubble wall velocity during the first-order phase transition \cite{vandeVis:2025plm}, 
etc. The measurements of cosmic microwave background anisotropies and Big Bang 
nucleosynthesis have inferred \cite{WMAP:2003ogi,Planck:2018vyg,Fields:2019pfx}
\begin{eqnarray}
\eta_B
=(8.8\pm 0.6)\times 10^{-11}.
\end{eqnarray}
We thus claim that the observed BAU can be explained in the SM4 with reasonable
inputs of parameters.

\section{Conclusion}
\label{sec5}

We have continued our endeavors to investigate the important dynamics inherent in the SM4 from 
various experimental and theoretical aspects. The SM4 with the specific fermion mass hierarchy 
survives the current experimental bounds \cite{Li:2023fim,Li:2024xnl}, breaks the 
electroweak symmetry by means of heavy fermion condensates 
\cite{Holdom:1986rn,Bardeen:1989ds,Hill:1990ge,Elliott:1992xg,Hung:2010xh,Mimura:2012vw}, 
achieves the first-order EWPT 
\cite{Ham:2004xh,Carena:2004ha,Fok:2008yg,Kikukawa:2009mu} through 
additional interactions between Higgs bosons and bound-state scalars of fourth generation 
quarks \cite{Li:2025bon}, and offers a viable CPV source for the BAU \cite{Hou:2008xd}. 
We then further demonstrated quantitatively that the prediction for the BAU sourced by fourth 
generation leptons accommodates the observed baryon-over-entropy ratio $\eta_B$. Instead of 
engaging the cumbersome bubble and transport dynamics imbedded in the first-order EWPT, 
we resorted to the effective theory approach which has been thoroughly 
studied, constructing the dimension-6 effective operators induced by 
fourth generation quarks with the CPV source from the $4\times 4$ CKM matrix. The CPV 
strength was shown to be proportional to the Yukawa couplings of fermions involved in the 
effective operators. The necessary inputs of the fourth generation fermion masses and the 
$4\times 4$ CKM matrix elements were obtained in the dispersive analyses on heavy quark decays 
and neutral quark state mixing. It is why the signs and the CPV strengths of the 
effective operators could be fixed unambiguously in our formalism. 

It has been argued that the transport equations for fourth generations leptons are similar
to those for third generation leptons, as the same relaxation rates and Yukawa rates 
are assumed. The solutions for their net number densities then share the similar behavior, 
and their contributions to the BAU are proportional to the CPV strengths of the corresponding 
effective operators. Hence, it is straightforward to extract the predictions for the BAU, 
given the same benchmark parameters, by scaling the $\tau$ lepton sourced results 
available in the literature. The $\tau$ sourced BAU in our SM4 setup is down by the 
new physics scale designated via the RG evolutions of the Yukawa couplings. The role of 
the $\tau$ source is replaced by the $\tau'$ and $\nu'$ sources, and the predicted ratio 
$\eta_B\approx 10^{-10}$ is in consistency with the observed value. Our series of 
works manifests that fourth generation quarks play the major role (through their condensates
and bound states) for accomplishing the electroweak 
symmetry breaking and the first-order EWPT, while fourth generation leptons with the CPV source 
from the $4\times 4$ CKM matrix are responsible for the realization of the BAU.

We have also taken this chance to revisit the dispersive constraints on the neutrino 
masses and the PMNS matrix elements in the presence of fourth generation leptons. It 
was found that a heavy fourth generation demands the almost exact unitarity of the 
$3\times 3$ PMNS matrix. Solving the dispersive constraints in an alternative way, we 
corroborate that the known inputs of the mass-squared differences $\Delta m^2_{21}$ and 
$\Delta m^2_{32}$ lead to the mixing angles and $CP$ phase in the $3\times 3$ PMNS matrix
$\theta_{12}\approx 34^\circ$, $\theta_{23}\approx 47^\circ$, 
$\theta_{13}\approx 5^\circ$ and $\delta\approx 200^\circ$. 
These results are insensitive to the variation of the lightest neutrino mass $m_1$, and 
close to the data for the normal ordering scenario. The distinct mixing patters in the 
quark and lepton sectors is attributed to the different quark and lepton mass spectra. 
The angle $\theta_{23}$ is preferred to be in the second octant. The 
consistencies of our SM4 setup in various aspects encourage the search for heavy fourth 
generation fermions at the (high-luminosity) large hadron collider or a muon collider.

\appendix
\section{Pseudoscalar penguin}
\label{appA}

We construct the pseudoscalar penguin operator, which will facilitate the derivation 
of the dimension-6 effective operator with a CPV source. The computation is similar to that in
\cite{Morozumi:2018cnc}, where a heavy vector quark is integrated out to produce effective 
operators in terms of SM fields. The Feynman rules for charged-scalar-quark vertices in the 
$R_\xi$ gauge with $\xi=0$ are written as
\begin{eqnarray}
iV_{ud}^*\left(g_u\frac{1+\gamma_5}{2}-g_d\frac{1-\gamma_5}{2}\right),\;\;\;\;
iV_{ud}\left(g_u\frac{1-\gamma_5}{2}-g_d\frac{1+\gamma_5}{2}\right),
\end{eqnarray}
for an incoming up-type quark and an outgoing down-type quark, and for an incoming down-type 
quark and an outgoing up-type quark, respectively, where $V_{ud}$ is the corresponding
CKM matrix element, and $g_u$ ($g_d$) is the Yukawa coupling of the up-type (down-type) 
quark. The pseudoscalar-fermion vertex reads
\begin{eqnarray}
-\sqrt{2}g_fI_{3f}\gamma_5,
\end{eqnarray}
$I_{3f}$ being the third component of the fermion isospin.

Figure~\ref{fig3} contains the relevant diagrams for the $t'\to t$ transition, where the
incoming $t'$ (outgoing $t$) quark carries the momentum $p_1$ ($p_2$). The vertex correction 
in Fig.~\ref{fig3}(a) consists of a charged scalar loop with a pseudoscalar boson attaching 
the internal $b'$ quark line. It gives the loop integral
\begin{eqnarray}
P_V&=&-\sqrt{2}V_{t'b'}^*V_{tb'}g_{b'}I_{3b'}
\int\frac{d^4 l}{(2\pi)^4}\left(-ig_{b'}\frac{1+\gamma_5}{2}\right)
\frac{i(\not p_2-\not l +m_{b'})}{(p_2-l)^2-m_{b'}^2}\gamma_5\nonumber\\
& &\hspace{3.0cm}\times\frac{i(\not p_1-\not l + m_{b'})}{(p_1-l)^2-m_{b'}^2}
i\left(g_{t'}\frac{1+\gamma_5}{2}-g_{b'}\frac{1-\gamma_5}{2}\right)\frac{i}{l^2}.
\label{bb1}
\end{eqnarray} 
We keep only the dominant term $g_{t'}(1+\gamma_5)/2$ in the parentheses in
Eq.~(\ref{bb1}) because of $g_{t'}>g_{b'}$, arriving at 
\begin{eqnarray}
P_V&=&-\frac{i}{\sqrt{2}}V_{t'b'}^*V_{tb'}I_{3b'}g_{b'}^2g_{t'}\int\frac{d^4 l}{(2\pi)^4}
\frac{(\not p_2-\not l)(\not p_1-\not l) -m_{b'}^2}{[(p_2-l)^2-m_{b'}^2][(p_1-l)^2-m_{b'}^2]l^2}
(1+\gamma_5).\label{b28}
\end{eqnarray} 
The numerator can be organized into 
\begin{eqnarray}
(\not p_2-\not l)(\not p_1-\not l) -m_{b'}^2
=(\not p_1-\not l-\not q)(\not p_1-\not l) 
=[(p_1-l)^2-m_{b'}^2]-\not q(\not p_1-\not l),\label{bb6}
\end{eqnarray}
with the momentum transfer $q=p_1-p_2$. The first term $(p_1-l)^2-m_{b'}^2$ cancels the 
one in the denominator, yielding
\begin{eqnarray}
-\frac{i}{\sqrt{2}}V_{t'b'}^*V_{tb'} I_{3b'}g_{b'}^2g_{t'}\int\frac{d^4 l}{(2\pi)^4}
\frac{1+\gamma_5}{[(p_2-l)^2-m_{b'}^2]l^2}.\label{b29}
\end{eqnarray}


The self-energy correction associated with the $t'\to t$ transition in Fig~\ref{fig3}(b)
is written as
\begin{eqnarray}
P_S&=&-\sqrt{2}V_{t'b'}^*V_{tb'} g_{t'}I_{3t'}
\int\frac{d^4 l}{(2\pi)^4}\left(-ig_{b'}\frac{1+\gamma_5}{2}\right)
\frac{i(\not p_2-\not l +m_{b'})}{(p_2-l)^2-m_{b'}^2}\nonumber\\
& &\hspace{3.0cm}\times\left(ig_{t'}\frac{1+\gamma_5}{2}\right)
\frac{i(\not p_2+ m_{t'})}{p_2^2-m_{t'}^2}\gamma_5\frac{i}{l^2}.
\end{eqnarray}
To simplify the calculation and grasp the important contribution, we drop $p_2$ relative 
to the largest mass $m_{t'}$, attaining
\begin{eqnarray}
P_S&=&-\sqrt{2}V_{t'b'}^*V_{tb'} g_{t'}I_{3t'}\frac{-ig_{b'}g_{t'}}{2}
\int\frac{d^4 l}{(2\pi)^4}\frac{m_{b'}}{[(p_2-l)^2-m_{b'}^2]l^2}
(1+\gamma_5)\frac{-1}{m_{t'}}\gamma_5\nonumber\\
&=&-\frac{i}{\sqrt{2}}V_{t'b'}^*V_{tb'} I_{3t'}g_{b'}^2g_{t'}
\int\frac{d^4 l}{(2\pi)^4}\frac{1+\gamma_5}{[(p_2-l)^2-m_{b'}^2]l^2}.
\label{bb3}
\end{eqnarray} 
The above integral is identical to Eq.~(\ref{b29}) except the isospin component 
$I_{3t'}=-I_{3b'}$, so they cancel each other exactly. It indicates that the 
pseudoscalar penguin is UV finite, and the difference between the vertex and self-energy 
corrections diminishes as $q\to 0$.

We turn to the net contribution from the second term on the right-hand side of
Eq.~(\ref{bb6}). The numerator becomes, after the variable change $l\to l+xp_1+yp_2$ 
and the Feynman parametrization,
\begin{eqnarray}
& &-\not q(\not p_1-\not l)\to x\not q\not p_1-q^2,
\label{nu1}
\end{eqnarray}
where the term linear in the loop momentum $l$ has been removed. The first term, leading to 
an integrand with an odd power of the loop momentum in Fig.~\ref{fig2}(b), does not 
contribute. Note that the momentum $p_1$ will be treated as a loop momentum in the
evaluation of Fig.~\ref{fig2}(b). Equation~(\ref{nu1}) then appears in the desired form; 
the factor $q^2$ cancels the propagator of the virtual pseudoscalar, which is proportional 
to $1/q^2$, producing a local operator. The Feynman parametrization recasts the 
denominator into
\begin{eqnarray}
& &l^2-(xp_1+yp_2)^2+x(p_1^2-m_{b'}^2)+y(p_2^2-m_{b'}^2)\nonumber\\
&=&l^2+x(1-x-y)p_1^2+y(1-x-y)p_2^2-(x+y)m_{b'}^2+xyq^2.\label{den}
\end{eqnarray}
The loop integral is dominated by the configuration, where the momentum $p_1$ is roughly 
on-shell with $p_1^2\approx m_{b'}^2$, and splits into $p_2$ and $q$ with smaller invariants 
$p_2^2$ and $q^2$. Equation~(\ref{den}) is thus approximated by
\begin{eqnarray}
l^2-[(1+x)(x+y)-x]m_{b'}^2.
\end{eqnarray}


We have the sum 
\begin{eqnarray}
P_V+P_S&=&-\frac{i}{\sqrt{2}}V_{t'b'}^*V_{tb'} I_{3b'}g_{b'}^2g_{t'}\int_0^1dx\int_0^{1-x}dy
\int\frac{d^4 l}{(2\pi)^4}\frac{-q^2(1+\gamma_5)}{\{l^2-[(1+x)(x+y)-x]m_{b'}^2\}^3}
\nonumber\\&=&\frac{1}{192\sqrt{2} }V_{t'b'}^*V_{tb'} I_{3b'}g_{b'}^2g_{t'}
\frac{q^2}{m_{b'}^2}(1+\gamma_5).\label{bb4}
\end{eqnarray}
At last, we multiply the above expression by the virtual pseudoscalar propagator and the 
spinors of the initial- and final-state fermions, establishing the 
four-fermion operator
\begin{eqnarray}
P&=&\frac{1}{192\sqrt{2}}V_{t'b'}^*V_{tb'} I_{3b'}g_{b'}^2g_{t'}
\frac{q^2}{m_{b'}^2}\bar t(1+\gamma_5)t'\frac{i}{q^2}(-\sqrt{2}g_fI_{3f})
\bar f\gamma_5f\nonumber\\
&=&-\frac{i}{192}V_{t'b'}^*V_{tb'} I_{3b'}I_{3f}\frac{g_fg_{b'}^2g_{t'}}{m_{b'}^2}
\bar t(1+\gamma_5)t'\bar f\gamma_5f.
\end{eqnarray}
It is learned that the pseudoscalar penguin can be constructed in the heavy quark limit, 
similar to the penguin operators in the conventional effective weak Hamiltonian. We 
emphasize that the above derivation does not apply, as the virtual pseudoscalar is replaced 
by a virtual scalar. In this case the mass squared term in Eq.~(\ref{b28}) would flip sign, 
such that one cannot have the exact cancellation of the $q$-independent pieces between the 
vertex and self-energy corrections.
 
It is trivial to obtain the self-energy correction associated with the $t\to t'$ 
transition,
\begin{eqnarray}
S&=&V_{tb}^*V_{t'b}\int\frac{d^4 l}{(2\pi)^4}ig_{t'}\frac{1-\gamma_5}{2}
\frac{i(\not p-\not l)}{(p-l)^2}ig_t\frac{1+\gamma_5}{2}\frac{i}{l^2}\nonumber\\
&=&\frac{i}{64\pi^2}V_{tb}^*V_{t'b} g_{t'}g_t\left(\ln\frac{\Lambda_s^2}{-p^2}+1\right)
\not p(1+\gamma_5),\label{bb5}
\end{eqnarray}  
with the $t$ quark momentum $p$. Because the full theory holds up to the electroweak 
symmetry restoration scale, we have bounded the loop momentum at $\Lambda_s$.

\section{$4\times 4$ PMNS matrix} 
\label{appB}

We review the dispersive constraints on the $4\times 4$ PMNS matrix, which were extracted 
from the mixing amplitudes associated with the neutral lepton pairs $\mathcal{L}^-\ell^+$ 
and $\mathcal{L}^+\ell^-$ \cite{Li:2024xnl}, $\mathcal{L}$ ($\ell$) being a massive (light) 
lepton. Our discussion covers the $\mu e$ and $\tau e$ systems, but not the 
$\tau\mu$ mixing; the mass ratio $m_{\tau}/m_\mu\sim O(10)$, much lower than 
$m_t/m_c\sim O(100)$, hints that the present formalism for a heavy-light system may not 
work for the $\tau\mu$ mixing. The $tc$ mixing was considered in Sec.~\ref{sec3}. It is not 
necessary to touch the $\tau' e$ mixing, because we have known that $U_{\tau'1}$, $U_{\tau'2}$ 
and $U_{\tau'3}$ diminish and $U_{\tau'4}\approx 1$ in the fourth row of the PMNS matrix. We 
will solve the dispersive constraints in a manner different from that in \cite{Li:2024xnl}. 
The same conclusion drawn in this appendix verifies the validity of the approximate 
solutions in \cite{Li:2024xnl}. Besides, the formalism presented here applies to the
$4\times 4$ CKM matrix for the quark mixing in Sec.~\ref{sec3} straightforwardly.

We decompose the mixing amplitude $\Pi(m_\mathcal{L}^2)$ into a sum over intermediate 
neutrino channels,
\begin{eqnarray}
\Pi(m_\mathcal{L}^2)&=& M(m_\mathcal{L}^2)-\frac{i}{2}\Gamma(m_\mathcal{L}^2)\nonumber\\
&\equiv& \sum_{i,j=1}^4\lambda_{i}\lambda_{j}\left[M_{ij}(m_\mathcal{L}^2)
-\frac{i}{2}\Gamma_{ij}(m_\mathcal{L}^2)\right],
\end{eqnarray} 
where $\lambda_i\equiv U^*_{\mathcal{L} i}U_{\ell i}$ is the product of the $4\times 4$ PMNS 
matrix elements with $i$ referring to the $i$-th generation neutrino. It has been illustrated 
that the electroweak symmetry is restored above a scale $\Lambda_s$ in our SM4 setup
\cite{Li:2025bon}, and that the mixing phenomenon disappears as 
$m_\mathcal{L}>\Lambda_s$; all internal particles become massless in the symmetric 
phase, such that the summation over all intermediate channels is strongly suppressed by the 
unitarity of the PMNS matrix. Small finite contributions to the real part 
$M_{ij}(m_\mathcal{L}^2)$ begin only at the three-loop level in the symmetric 
phase \cite{Li:2024awx}. The corresponding dispersion relation is thus expressed as
\begin{eqnarray}
M(m_\mathcal{L}^2)=\frac{1}{2\pi}\int^{\Lambda_s^2} dm^2
\frac{\Gamma(m^2)}{m_\mathcal{L}^2-m^2}\approx 0,\label{dis}
\end{eqnarray}
for $m_\mathcal{L}>\Lambda_s$, where the upper bound of the integration variable 
$m^2$ is set to $\Lambda_s^2$.

To obey Eq.~(\ref{dis}) for arbitrary $m_\mathcal{L}>\Lambda_s$, some conditions must 
be met by the PMNS matrix elements appearing in the imaginary part $\Gamma(m^2)$. We quote the 
asymptotic behavior of the component $\Gamma_{ij}(m^2)$ for $m<\Lambda_s$ 
\cite{Cheng:1982hq,Buras:1984pq}
\begin{eqnarray}
\Gamma_{ij}(m^2)\approx \Gamma^{(1)}_{ij}m^2+\Gamma^{(0)}_{ij}
+\frac{\Gamma^{(-1)}_{ij}}{m^2}+\cdots,\label{exp}
\end{eqnarray}
where the coefficients $\Gamma^{(n)}_{ij}$, $n=1,0,-1$, have been presented in \cite{Li:2024xnl}.
These terms give contributions to the dispersive integral in Eq.~(\ref{dis}), which scale 
like $\Lambda_s^4$, $\Lambda_s^2$ and $\ln\Lambda_s^2$ for $n=1,0$ and $-1$, respectively. 
Since the upper bound $\Lambda_s$ represents an order-of-magnitude concept, instead of a definite 
value, it is unlikely that the above huge contributions happen to cancel among themselves. 
Therefore, the finiteness of the dispersive integral must be fulfilled by vanishing the sum
of the coefficients
\begin{eqnarray}
\sum_{i,j=1}^4\lambda_{i}\lambda_{j} \Gamma_{ij}^{(n)}\approx 0,\;\;\;\;n=1,0,-1.\label{cons}
\end{eqnarray}

Once the conditions in Eq.~(\ref{cons}) are conformed, we rewrite the dispersive integral 
in Eq.~(\ref{dis}) as
\begin{eqnarray}
\int^{\Lambda_s^2} dm^2\frac{\Gamma(m^2)}{m_\mathcal{L}^2-m^2}\approx
\frac{1}{m_\mathcal{L}^2}\sum_{i,j=1}^4\lambda_{i}\lambda_{j}g_{ij},\label{cong}
\end{eqnarray}
with the factors
\begin{eqnarray}
g_{ij}\equiv\int_{t_{ij}}^\infty dm^2\left[\Gamma_{ij}(m^2)-
\Gamma_{ij}^{(1)}m^2-\Gamma^{(0)}_{ij}-\frac{\Gamma^{(-1)}_{ij}}{m^2}\right],\label{gi}
\end{eqnarray}
$t_{ij}=(m_i+m_j)^2$ being the threshold for one of the channel. The integrand in the square 
brackets decreases like $1/m^4$, so the upper bound of $m^2$ in Eq.~(\ref{gi}) can be pushed 
to infinity safely. The approximation $1/(m_\mathcal{L}^2-m^2)\approx 1/m_\mathcal{L}^2$ has 
been made for large $m_\mathcal{L}>\Lambda_s$, because the integral receives contributions 
only from finite $m<\Lambda_s$. We place the final condition, 
\begin{eqnarray}
\sum_{i,j=1}^4\lambda_{i}\lambda_{j}g_{ij}\approx 0,\label{gij}
\end{eqnarray}
to guarantee an almost nil dispersive integral. That is, Eqs.~(\ref{cons}) and (\ref{gij}) 
constitute a solution to the integral equation (\ref{dis}). 

We eliminate $\lambda_4=-\lambda_1-\lambda_2-\lambda_3$ using the unitarity condition, 
and define the ratios of the PMNS matrix elements,
\begin{eqnarray}
r_1\equiv\frac{U^*_{\mathcal{L} 1}U_{e1}}{U^*_{\mathcal{L} 2}U_{e2}}\equiv u_1+iv_1,\;\;\;\;
r_3\equiv\frac{U^*_{\mathcal{L} 3}U_{e3}}{U^*_{\mathcal{L} 2}U_{e2}}\equiv u_3+iv_3,
\end{eqnarray}
where $\mathcal{L}$ stands for either $\mu$ or $\tau$, and the unknowns $u_{1,3}$ and 
$v_{1,3}$ will be constrained below. The $n=1,0$ and $-1$ conditions are then formulated 
as \cite{Li:2024xnl}
\begin{eqnarray}
& &\left(r_1\frac{m_4^2-m_1^2}{m_W^2-m_1^2}+\frac{m_4^2-m_2^2}{m_W^2-m_2^2}
+r_3\frac{m_4^2-m_3^2}{m_W^2-m_3^2}\right)^2
\approx 0,\label{cs1}\\
& &\left(r_1\frac{m_4^2-m_1^2}{m_W^2-m_1^2}+\frac{m_4^2-m_2^2}{m_W^2-m_2^2}
+r_3\frac{m_4^2-m_3^2}{m_W^2-m_3^2}\right)\nonumber\\
& &\times\left(r_1\frac{m_4^2-m_1^2}{m_W^2-m_1^2}P_1
+\frac{m_4^2-m_2^2}{m_W^2-m_2^2}P_2+r_3\frac{m_4^2-m_3^2}{m_W^2-m_3^2}P_3\right)
\approx 0,\label{cs2}\\
& &\left(r_1\frac{m_4^2-m_1^2}{m_W^2-m_1^2}+\frac{m_4^2-m_2^2}{m_W^2-m_2^2}
+r_3\frac{m_4^2-m_3^2}{m_W^2-m_3^2}\right)\nonumber\\
& &\times\left(r_1\frac{m_4^2-m_1^2}{m_W^2-m_1^2}Q_1
+\frac{m_4^2-m_2^2}{m_W^2-m_2^2}Q_2
+r_3\frac{m_4^2-m_3^2}{m_W^2-m_3^2}Q_3\right)\approx 0,\label{cs3}
\end{eqnarray}
respectively, with the functions
\begin{eqnarray}
P_i&=&\frac{m_W^2(m_4^2+m_i^2)-m_4^2m_i^2}{m_4^4}\nonumber\\
Q_i&=&\frac{2m_W^4(m_4^2+m_i^2)+m_4^2m_i^2(m_4^2+m_i^2)-m_W^2(m_4^4+3m_4^2m_i^2+m_i^4)}{m_4^6}.
\label{pq}
\end{eqnarray}
The last condition in Eq.~(\ref{gij}) turns into \cite{Li:2024xnl}
\begin{eqnarray}
& &-(20m_W^4-28m_W^2m_4^2+7m_4^4)G_0
+(16m_W^4-28m_4^2m_W^2+ 12m_4^4)G_1 \approx 0,\label{fs}\\
& &G_0=\left(\frac{r_1}{m_W^2-m_1^2}+\frac{1}{m_W^2-m_2^2}
+\frac{r_3}{m_W^2-m_3^2}\right)^2,\label{fs0}\\
& &G_1=\left(\frac{r_1}{m_W^2-m_1^2}+\frac{1}{m_W^2-m_2^2}
+\frac{r_3}{m_W^2-m_3^2}\right)\nonumber\\
& &\hspace{1.2cm}\times\left[\frac{r_1 m_1}{m_4(m_W^2-m_1^2)}
+\frac{m_2}{m_4(m_W^2-m_2^2)}+\frac{r_3 m_3}{m_4(m_W^2-m_3^2)}\right].\label{fs1}
\end{eqnarray}

Our alternative strategy to solve the coupled Eqs.~(\ref{cs1})-(\ref{cs3}) and (\ref{fs}) 
is detailed as follows. We first solve for $u_3$ and $v_3$ in terms of $u_1$ and $v_1$
from the real part of Eq.~(\ref{cs1}) 
\begin{eqnarray}
u_1\frac{m_4^2-m_1^2}{m_W^2-m_1^2}+\frac{m_4^2-m_2^2}{m_W^2-m_2^2}
+u_3\frac{m_4^2-m_3^2}{m_W^2-m_3^2}=
v_1\frac{m_4^2-m_1^2}{m_W^2-m_1^2}+v_3\frac{m_4^2-m_3^2}{m_W^2-m_3^2},\label{a1}
\end{eqnarray}
and the real part of Eq.(\ref{fs}). Inserting the expressions of $u_3$ and $v_3$ into
the rest of constraints, we minimize the deviations of the imaginary part of Eq.~(\ref{cs1}),
both real and imaginary parts of Eq.~(\ref{cs2}), and the imaginary part of Eq.~(\ref{fs})
by tuning $u_1$ and $v_1$. The coupled equations are fully respected in this sense. 
The difference between Eqs.~(\ref{cs3}) and (\ref{cs2}) are numerically negligible, 
so the former can be excluded. We assume a small first generation mass $m_1^2=10^{-6}$ eV$^2$ 
\cite{Li:2023ncg}, and input $m_2$ and $m_3$ from the measured mass-squared differences 
$\Delta m^2_{21} \equiv m^2_{2}-m^2_{1}= (7.50^{+0.22}_{-0.20})\times 10^{-5}$ eV$^2$ 
and $\Delta m^2_{31}\equiv m^2_{3}-m^2_{1}=(2.55^{+0.02}_{-0.03})\times 10^{-3}$ eV$^2$ in 
the normal-ordering scenario \cite{deSalas:2020pgw}. It has been demonstrated \cite{Li:2023ncg}
that the inverted mass ordering is not favored by the dispersive constraints.
The above parameters were updated in \cite{Tortola2024,Capozzi:2025wyn} with tiny changes, 
such as $\Delta m^2_{21} = (7.55^{+0.22}_{-0.20})\times 10^{-5}$ eV$^2$ \cite{Tortola2024}. 
The fourth generation neutrino mass and the $W$ boson mass take $m_{\nu'}=170$ GeV and 
$m_W=80.37$ GeV, respectively.


The tuning of $u_1$ and $v_1$ is proceeded in a way similar to that in \cite{Li:2023ncg}.
We first inspect the dependencies of the aforementioned four constraints on $u_1$ for $v_1=0$. 
It is seen that the first three have a single root $u_1\approx 0.97$, while the
last one reveals two roots $u_1\approx -0.97$ and $u_1\approx -0.83$, which arise from
Eq.~(\ref{fs1}),
\begin{eqnarray}
& &\left(\frac{u_1}{m_W^2-m_1^2}+\frac{1}{m_W^2-m_2^2}+\frac{u_3}{m_W^2-m_3^2}\right)\nonumber\\
& &\times \left(u_1\frac{m_4^2-m_1^2}{m_W^2-m_1^2}+\frac{m_4^2-m_2^2}{m_W^2-m_2^2}
+u_3\frac{m_4^2-m_3^2}{m_W^2-m_3^2}\right)=0.\label{c1}
\end{eqnarray}
When Eq.~(\ref{cs1}) is satisfied, the magnitude of Eq.~(\ref{fs0}) diminishes, for 
Eqs.~(\ref{cs1}) and (\ref{fs0}) differ only by $O(m_i^2/m_{4,W}^2)$. Namely, Eq.~(\ref{fs}) 
is governed by the second term, which should vanish as shown in Eq.~(\ref{c1}). It 
describes a parabola in $u_1$, explaining the existence of the two roots. 
It is encouraging that $u_1\approx -0.97$ and $u_1\approx -0.83$ are close 
to the real parts of $r_1$ associated with the $\mu e$ and $\tau e$ mixings,
respectively. To meet the four constraints for the $\mu e$ and $\tau e$ mixings
simultaneously, $v_1$, which is related to the $CP$ phase of the PMNS matrix, must
take a finite value; for a finite $v_1$, the first three constraints develop an
additional root of $u_1$ around $u_1\approx -0.83$. We increase $v_1$ gradually until 
this additional root of $u_1$ coincides with the minimal deviation of the fourth condition 
from zero. This best-fit $v_1$ is found to be $v_1\approx -0.04$, corresponding to 
which the two roots of $u_1$ shift a bit to $-0.98$ and $-0.84$.   

As pointed out in \cite{Li:2024xnl}, $v_1$ is determined up to a sign; the right-hand 
side of Eq.~(\ref{a1}) can have a minus sign, which also minimizes the real part of 
Eq.~(\ref{cs1}). We then arrive at the solutions of $r_1$ for the $\mu e$ and $\tau e$ 
mixings,
\begin{eqnarray}
\frac{U^*_{\mu 1}U_{e1}}{U^*_{\mu 2}U_{e2}}=-0.84-0.04i,\;\;\;\;
\frac{U^*_{\tau 1}U_{e1}}{U^*_{\tau 2}U_{e2}}=-0.98+0.04i,\label{ns1}
\end{eqnarray}
where the sign of $v_1$ has been assigned appropriately. Equation~(\ref{ns1}) is insensitive 
to the choices of the light neutrino mass $m_1$, as long as it is sufficiently lower than 
$m_2$. Equation~(\ref{ns1}) is almost identical to $-0.83-0.04i$ and $-0.98+0.04i$ 
derived in \cite{Li:2024xnl}. We mention that the solutions for $r_{1,3}$ well 
respect the unitarity of the $3\times 3$ PMNS matrix. This is obvious from the constraints 
in Eqs.~(\ref{cs1})-(\ref{cs3}) and (\ref{fs}), which require $u_1+u_3\approx -1$ and 
$v_1+v_3\approx 0$. For example, inserting $u_1=-0.84$ and $v_1=-0.04$ into the expressions 
for $u_3$ and $v_4$, we get $u_3=-0.26$ and $v_3=0.04$. 

The above formalism certainly extends to the quark mixing, such as the $c\bar u$-$\bar c u$ 
and $t\bar u$-$\bar t u$ mixings with $d$, $s$, $b$ and $b'$ quarks in the intermediate 
channels. The corresponding dispersive constraints are copied from those for the lepton 
mixing by substituting the quark masses $m_d$, $m_s$, $m_b$ and $m_{b'}$
for $m_1$, $m_2$, $m_3$ and $m_4$, respectively. As remarked in Sec.~\ref{sec3}, different 
solutions for the quark and lepton mixings come from the disparate 
mass spectra of quarks and leptons. The goal of the dispersive analysis on the 
$4\times 4$ CKM matrix is to estimate the fourth column and fourth row elements, which 
may not diminish, and are essential for assessing the CPV strength of the dimension-6 effective 
operators.


\acknowledgments

We thank K.F. Chen, Y.T. Chien, C.S. Chu, S.Y. Ho, W.S. Hou, Y.J. Lin, S.M. Wang, M.R. Wu,
Y.L. Wu, X.B. Yuan and X.M. Zhang for fruitful discussions. We also thank the Yukawa Institute 
for Theoretical Physics at Kyoto University, where part of the work was completed during 
“Progress of Theoretical Bootstrap”. This work was supported by National Science and 
Technology Council of the Republic of China under Grant No. NSTC-113-2112-M-001-024-MY3.



\end{document}